\documentclass[smallextended]{svjour3}
\usepackage[T1]{fontenc}
\usepackage[latin9]{inputenc}
\usepackage{amsmath}
\usepackage{graphicx}
\usepackage{epstopdf}
\usepackage{xcolor}
\usepackage{babel}
\usepackage[unicode=true,pdfusetitle,
 bookmarks=true,bookmarksnumbered=false,bookmarksopen=false,
 breaklinks=true,pdfborder={0 0 0},pdfborderstyle={},backref=false,colorlinks=true]
 {hyperref}

\newcommand{\orcid}[1]{\href{https://orcid.org/#1}{\includegraphics[height = 2ex]{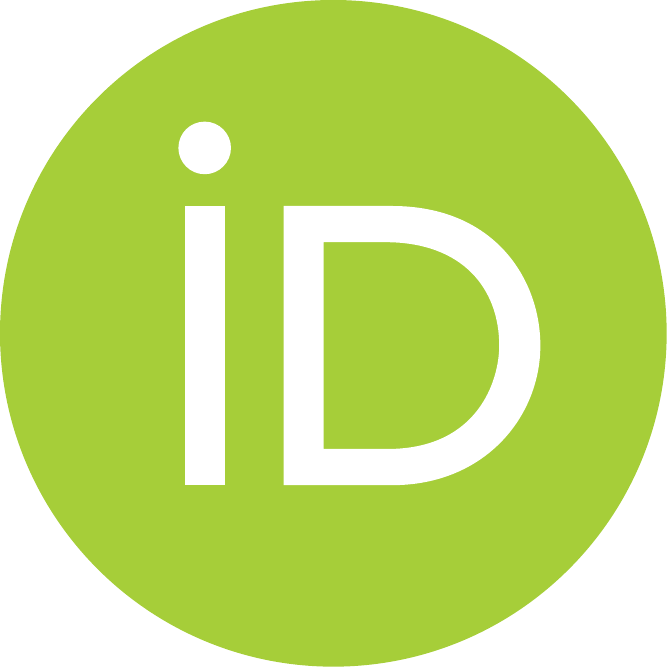}}}

\begin{document}\sloppy

\title{Quantum Divide and Compute: Exploring The Effect of Different Noise Sources}

\author{Thomas~Ayral${}^{*}$ \orcid{0000-0003-0960-4065}, Fran\c{c}ois-Marie~Le R\'egent \orcid{0000-0002-5229-7155}, Zain~Saleem \orcid{0000-0002-8182-2764}, Yuri~Alexeev \orcid{0000-0001-5066-2254},
Martin~Suchara\orcid{0000-0001-8808-1367}}
\authorrunning{T. Ayral, F.-M. Le R\'egent, Z. Saleem, Y. Alexeev, M. Suchara}
\institute{Thomas Ayral, \at Atos Quantum Laboratory, Les Clayes-sous-Bois, France, 
Corresponding author (\href{mailto:thomas.ayral@atos.net}{thomas.ayral@atos.net}, +33 1 30 80 70 00)
\and 
F.M Le R\'egent  \at Atos Quantum Laboratory, Les Clayes-sous-Bois, France and  Ecole Polytechnique, Palaiseau, France,
\and 
Zain Saleem, Yuri Alexeev,
Martin Suchara \at 
Argonne National Laboratory, Lemont, Illinois, United States of America}

\maketitle

\begin{abstract}
Our recent work~\cite{Ayral2020} showed the first implementation of the Quantum Divide and Compute (QDC) method, which allows to break quantum circuits into smaller fragments with fewer qubits and shallower depth. QDC can thus deal with the limited number of qubits and short coherence times of noisy, intermediate-scale quantum processors. This article investigates the impact of different noise sources---readout error, gate error and decoherence---on the success probability of the QDC procedure. We perform detailed noise modeling on the Atos Quantum Learning Machine, allowing us to understand tradeoffs and formulate recommendations about which hardware noise sources should be preferentially optimized. We describe in detail the noise models we used to reproduce experimental runs on IBM's Johannesburg processor. This work also includes a detailed derivation of the equations used in the QDC procedure to compute the output distribution of the original quantum circuit from the output distribution of its fragments. Finally, we analyze the computational complexity of the QDC method for the circuit under study via tensor-network considerations, and elaborate on the relation the QDC method with tensor-network simulation methods.


\end{abstract}

\section{Introduction}

The advent of Noisy Intermediate Scale Quantum (NISQ) technologies~\cite{preskill2018} makes multiqubit processors with modest but increasing numbers of qubits available. Google, IBM, and Intel have recently announced quantum computers with 72, 65, and 49 qubits, respectively~\cite{Bristlecone_72,IBM_50,Intel_49}; and new systems with 50 to 200 qubits are expected to be commercially available in the next few years.
However, our ability to use the hardware to solve interesting problems is lagging.
Solving practical computational problems typically requires evaluating quantum circuits with many hundreds or even thousands of qubits, exceeding the size of the devices. In addition, large gate errors and short qubit coherence times prevent accurate evaluations of deep circuits.

Despite the remarkable progress in manufacturing and controlling these small multiqubit systems, building hardware with sufficiently high number of high-fidelity qubits remains an extremely challenging task. Engineering challenges worsen as the systems scale and are inherent for all major qubit technologies, including superconducting qubits (errors due to Josephson junction defects and spurious microwave resonances~\cite{scaling_superconducting}), ion traps (susceptibility to noise and difficulty to address individual ions~\cite{Monroe1164}), neutral atoms (motion of the atoms inside the lattice~\cite{Saffman_neutral_atoms}), and quantum dots (difficulty to entangle multiple qubits~\cite{entanglement_defect_spins,Heralded_entanglement}).

Successfully solving practical computational problems can be achieved only by developing techniques that can simultaneously map large problems onto small qubit systems and mitigate the effects of noise. The Quantum Divide and Compute (QDC) approach is one such technique.
In this approach, we divide a large and potentially deep quantum circuit to suit the number of qubits and coherence times available in the current quantum hardware.
We then perform the computations on the subcircuits obtained by this division on a quantum processor, and we finally recombine our output results to obtain the output of the original circuit.
This allows us to compute the outputs of quantum circuits that are too deep or too wide to be run on existing small-scale quantum processors.

There has been some previous work related to this approach. Bravyi et al.~\cite{Bravyi2016} showed that a quantum circuit on $n + k$ qubits can be simulated by sparse circuits on $n$ qubits and exponential classical processing that takes time $2^{O(k)}poly(n)$.
A more general approach that allows fragmenting larger quantum circuits into smaller subcircuits was introduced in~\cite{Peng2019}. In this work, tensor-network techniques were used to show how to decompose a circuit with a large quantum volume~\cite{Cross2019} into smaller subcircuits with quantum volumes compatible with NISQ devices. The classical computing overhead of the circuit fragmenting techniques was reduced in~\cite{perlin2020quantum}, and maximum likelihood tomography was applied on top of the circuit fragmentation to ensure that the reconstructed probability distributions are strictly non-negative and normalized.
This work also showed, with the help of classical simulations, that the QDC strategy, when combined with maximum likelihood tomography, can estimate the output of a clustered circuit with higher fidelity than the full circuit execution. In~\cite{Tang_CutQC}, a method was introduced to locate the optimal location of the cut (the location where the circuit should be fragmented). The QDC strategy was applied to commonly known circuits in quantum computing such as supremacy circuits, Grover and Bernstein-Vazirani circuits, and was shown to achieve a high quantum circuit evaluation fidelity.

The ultimate test for the quantum computing field---the ability to use controlled quantum systems to perform tasks surpassing what can be done using classical computers, also called quantum supremacy~\cite{1203.5813}---has received considerable attention from  both the scientific community and the general public. The largest classical supercomputers are capable of reliably simulating quantum systems with approximately 50 qubits~\cite{Alexeev_IntelQS,Haner:2017:PSQ:3126908.3126947}, and there is evidence that devices with more than 50 qubits may be able to demonstrate quantum supremacy even in the presence of noise~\cite{Boixo_supremacy}. 
While quantum supremacy is not one of the goals of this work, the developed techniques will allow increasing the size of circuits that can be evaluated on quantum hardware as well as on quantum simulators run on classical hardware~\cite{Qiskit,1601.07195,Atos,ProjectQ} by a constant factor. Consequently, it will be possible to evaluate quantum circuits with hundreds of qubits and use quantum algorithms to solve problems larger than ever before.

Circuit cutting naturally complements variational quantum-classical algorithms such as the Variational Quantum Eigensolver (VQE)~\cite{McClean_2016,PhysRevLett.110.090501} and the Quantum Approximate Optimization Algorithm (QAOA)~\cite{Farhi_ApproximateOptimization}. These approaches have successfully produced suitable quantum circuits for optimization problems by combining shallow quantum circuits with classical processing; and they allow some control over the width, depth, and connectivity of the circuits. However, the quality of the approximate solutions produced by VQE and QAOA decreases as the width and depth of their circuits decreases, and solving most interesting problems still requires hundreds of qubits~\cite{PhysRevA.92.042303,1812.07589}.

Circuit cutting offers numerous benefits. First, the technique does not compromise the quality of the solution as the size of the subcircuits decreases (overhead may scale exponentially with the number of cuts, however).
Second, the technique can be applied to any sparsely connected quantum circuit, irrespective of the structure of the problem. Third, circuit cutting has a close relationship with tensor network quantum simulation techniques that are used to address scalability limitations due to memory requirements that grow exponentially with the size of the simulated systems. Fourth, circuit cutting can enhance the performance of existing quantum-classical variational approaches because it can increase the size of the subproblems tackled by the variational quantum eigensolver.

In this article, we follow up on our previous work on the topic~\cite{Ayral2020}: we start by giving a detailed derivation of the formula for the output reconstruction of the original circuit from the outputs of its fragments, and a description of the noise models we chose to reproduce the experimental results (Section \ref{methods}).
We quantify the performance of the QDC method by recalling our previous results~\cite{Ayral2020} on a 20-qubit IBM processor for different qubit counts and fragment sizes (Section \ref{overview}).
Then, in Section \ref{results}, based on noisy simulations, we quantify the differential influence of various noise sources such as readout error, gate error and decoherence on the success probability of the algorithm for different qubit counts and fragment sizes.
Finally, we discuss the classical complexity of the method, its relation to tensor-network simulation approaches, and its implications for homogeneous and heterogeneous quantum computing.

\section{Methods}\label{methods}

\subsection{Circuit cutting}\label{subsec:Circuit-cutting}

An algorithm that allows circuit cutting was first described in~\cite{Peng2019}. In this section, we provide a self-contained derivation that allows to compute the probability distribution of a circuit that has been fragmented into several smaller disconnected pieces. We first derive a formula that uses probability distributions of two fragments to obtain the probability distribution of the original circuit. We then generalize the formula for cases with more than two fragments.

\subsubsection{Two-fragment case: definitions}

\begin{figure}
\begin{centering}
\includegraphics[width=1.0\columnwidth]{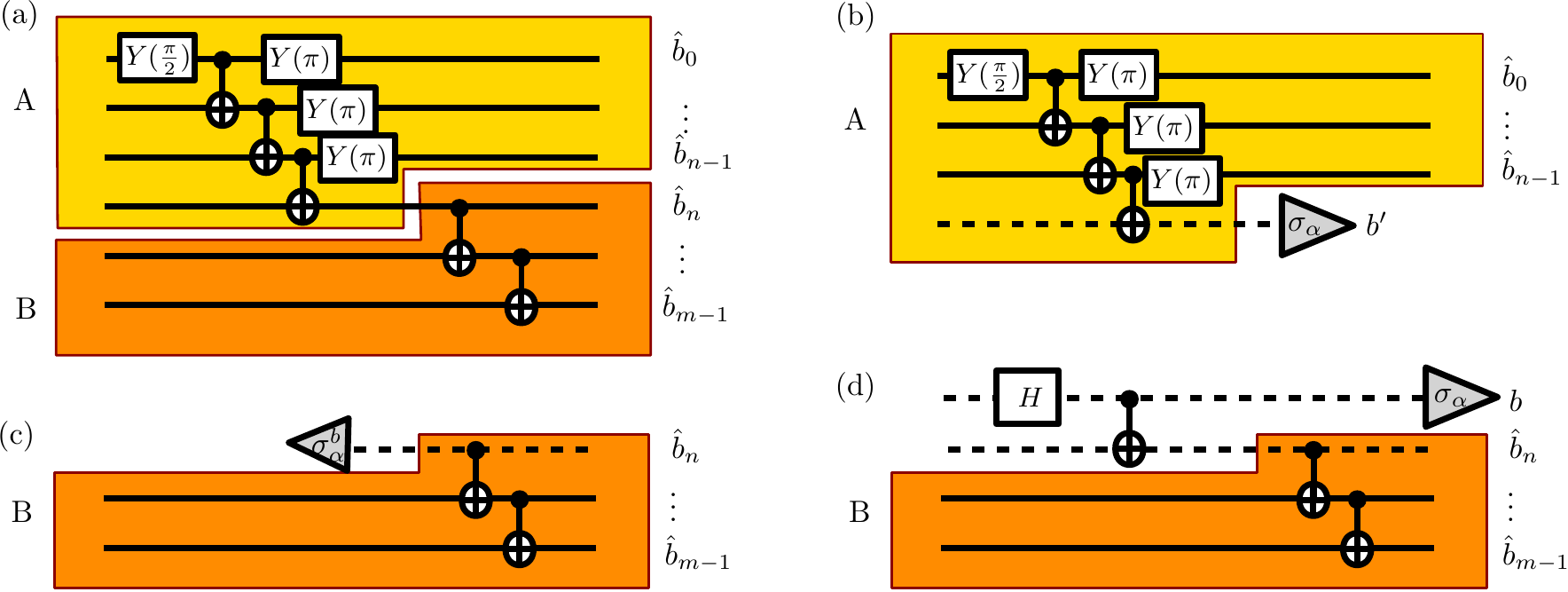}
\par\end{centering}
\caption{Cutting sketch in the two-fragment case. (a) Original circuit, (b) Upper fragment of the circuit, (c)
Lower fragment of the circuit, (d) Lower fragment in its Bell-state variant \label{fig:Cutting-sketch}}
\end{figure}

Let us consider a $m$-qubit circuit described as the following composition
of operations:
\[
\mathcal{O}=\mathcal{O}_{A}^{\mathrm{a}}\circ\mathcal{O}_{B}^{\mathrm{a}}\circ\mathcal{O}_{A}^{\mathrm{b}}\circ\mathcal{O}_{B}^{\mathrm{b}}
\]

where the support of superoperators $\mathcal{O}_{A}^{\mathrm{b}}$
and $\mathcal{O}_{B}^{\mathrm{b}}$ is a bipartition of the qubits;
similarly, the support of $\mathcal{O}_{A}^{\mathrm{a}}$ and $\mathcal{O}_{B}^{\mathrm{a}}$
is a bipartition such that the two ``a'' (for ``after'') sets
differ from the ``b'' (for ``before'') sets by one qubit. Without
loss of generality, one can assume that up to a relabeling, the support
of $\mathcal{O}_{A}^{\mathrm{b}}$ is $q_{0},\dots q_{n}$ and that
of $\mathcal{O}_{B}^{\mathrm{b}}$ is $q_{n+1},\dots q_{m-1}$, and
the ``a'' supports, $q_{0},\dots q_{n-1}$ and $q_{n},\dots q_{m-1}$
(see Fig. \ref{fig:Cutting-sketch}(a)).

The final state of the circuit is given by the density matrix:
\begin{align*}
\rho & =\mathrm{\mathcal{O}}(\rho_{0})=\mathcal{O}_{A}^{\mathrm{a}}\circ\mathcal{O}_{B}^{\mathrm{a}}\circ\mathcal{O}_{A}^{\mathrm{b}}\circ\mathcal{O}_{B}^{\mathrm{b}}(\rho_{0})
\end{align*}
where $\rho_0$ is the initial density matrix. The probability of measuring a state $i$ with binary representation
$i=(\hat{b}_{0}(i),\dots \hat{b}_{m-1}(i))$ is given by 
\begin{equation}
p(i)=\mathrm{Tr}\left[\Pi_{i}\cdot\rho\right]\label{eq:state_probability}
\end{equation}

where $\Pi_{i}$ is the projector on state $i$ ($i=0\dots2^{m}$).
It can be expressed as $\Pi_{i}=|i\rangle\langle i|=\otimes_{k=0}^{m-1}|\hat{b}_{k}(i)\rangle\langle \hat{b}_{k}(i)|$,
where $\hat{b}_{k}(i)$ is the value of the $k$th bit of $i$. We note that
$\Pi_{i}^{\dagger}=\Pi_{i}$, and $\sum_{i}\Pi_{i}=\otimes_{k}\sum_{\hat{b}_{k}=0}^{1}|\hat{b}_{k}\rangle\langle \hat{b}_{k}|=I$.
Thus:
\begin{equation}
p(i)=\mathrm{Tr}\left[\Pi_{i}^{\dagger}\cdot\mathcal{O}_{A}^{\mathrm{a}}\circ\mathcal{O}_{B}^{\mathrm{a}}\circ\mathcal{O}_{A}^{\mathrm{b}}\circ\mathcal{O}_{B}^{\mathrm{b}}(\rho_{0})\right].\label{eq:prob_distrib}
\end{equation}

We now switch to a Pauli-basis representation (see Appendix \ref{sec:Pauli-basis-representation}
for a reminder). Using Eq. (\ref{eq:trace_scalar_relation}), we get

\begin{equation}
p(i)=2^{m}\langle\langle\Pi_{i}|\mathcal{R}_{A}^{\mathrm{a}}\mathcal{R}_{B}^{\mathrm{a}}\mathcal{R}_{A}^{\mathrm{b}}\mathcal{R}_{B}^{\mathrm{b}}|\rho_{0}\rangle\rangle\label{eq:prob_distrib_Pauli}
\end{equation}

where $\mathcal{R}_{A/B}^{\mathrm{a/b}}$ is the Pauli transfer matrix
(PTM) representation of superoperator $\mathcal{O}_{A/B}^{\mathrm{a/b}}$.

\subsubsection{Bipartite splitting formula}

\paragraph{Basic formula}

We now derive the splitting formula. Let us decompose the one-qubit
PTM representation of the identity superoperator as
\begin{equation}
\mathcal{R}_{I}=\sum_{\alpha=X,Y,Z}\sum_{bb'\in\{0,1\}}\tilde{\gamma}_{\alpha}^{bb'}|\sigma_{\alpha}^{b}\rangle\rangle\langle\langle\sigma_{\alpha}^{b'}|\label{eq:identity_decomp}
\end{equation}

where $|\sigma_{\alpha}^{b}\rangle\rangle$ are the (real) coordinates
in the Pauli basis of the density matrix corresponding to the $b$th
eigenvector $|\psi_{\alpha}^{b}\rangle$ of Pauli matrix $\sigma_{\alpha}$.
The $\tilde{\gamma}$ tensor is given by $\tilde{\gamma}_{X}^{bb'}=\tilde{\gamma}_{Y}^{bb'}=2\delta_{bb'}-1$
and $\tilde{\gamma}_{Z}^{bb'}=2\delta_{bb'}$. 

Inserting $\mathcal{R}_{I}$ (acting on qubit $q_{n}$) in the expression
for the probability, Eq. (\ref{eq:prob_distrib_Pauli}), we obtain
\begin{align*}
p(i) & =2^{m}\langle\langle\Pi_{i}|\underbrace{\mathcal{R}_{A}^{\mathrm{a}}}_{q_{0},\dots q_{n-1}}\underbrace{\mathcal{R}_{B}^{\mathrm{a}}}_{q_{n},\dots q_{m-1}}\underbrace{\mathcal{R}_{I}}_{q_{n}}\underbrace{\mathcal{R}_{A}^{\mathrm{b}}}_{q_{0},\dots q_{n}}\underbrace{\mathcal{R}_{B}^{\mathrm{b}}}_{q_{n+1}\dots q_{m-1}}|\rho_{0}\rangle\rangle\\
 & =2^{m}\sum_{\alpha=X,Y,Z}\sum_{bb'\in\{0,1\}}\tilde{\gamma}_{\alpha}^{bb'}
 \;\;\times\langle\langle\Pi_{i}|_{q_{0}\dots q_{n-1}}\langle\langle\Pi_{i}|_{q_{n}\dots q_{m-1}}\underbrace{\mathcal{R}_{A}^{\mathrm{a}}}_{q_{0}\dots q_{n-1}}\underbrace{\mathcal{R}_{B}^{\mathrm{a}}}_{q_{n}\dots q_{m-1}}|\sigma_{\alpha}^{b}\rangle\rangle_{q_{n}}\\
 & \;\;\times\langle\langle\sigma_{\alpha}^{b'}|_{q_{n}}\underbrace{\mathcal{R}_{A}^{\mathrm{b}}}_{q_{0}\dots q_{n-1}}\underbrace{\mathcal{R}_{B}^{\mathrm{b}}}_{q_{n+1}\dots q_{m-1}}|\rho_{0}\rangle\rangle_{q_{0}\dots q_{n}}|\rho_{0}\rangle\rangle_{q_{n+1}\dots q_{m-1}}\\
 & =2^{m}\sum_{\alpha=X,Y,Z}\sum_{bb'\in\{0,1\}}\tilde{\gamma}_{\alpha}^{bb'}2^{-n-1}p_{A}^{\alpha}(i_{|0\dots n-1};b')\;2^{-m+n}
  \;\;\;  p_{B}^{\alpha b}(i_{|n\dots m-1}).
\end{align*}

We thus obtain the final formula (with $i=(\hat{b}_{0}\dots \hat{b}_{m-1})$):

\begin{align}
 & p(\hat{b}_{0}\dots \hat{b}_{m-1})=\label{eq:final_formula}
 \frac{1}{2}\sum_{\alpha=X,Y,Z}\sum_{bb'\in\{0,1\}}\tilde{\gamma}_{\alpha}^{bb'}p_{A}^{\alpha}(\hat{b}_{0}\dots \hat{b}_{n-1};b')p_{B}^{\alpha b}(\hat{b}_{n}\dots \hat{b}_{m-1})
\end{align}

with
\begin{align}
 & p_{A}^{\alpha}(\hat{b}_{0}\dots \hat{b}_{n-1};b')\equiv
 2^{n+1}\langle\langle\Pi_{\hat{b}_{0}\dots \hat{b}_{n-1}}|\langle\langle\sigma_{\alpha}^{b'}|_{q_{n}}\mathcal{R}_{A}|\rho_{0}\rangle\rangle_{q_{0}\dots q_{n}}\label{eq:proba_A_frag} \\
 & p_{B}^{\alpha b}(\hat{b}_{n}\dots \hat{b}_{m-1})\equiv
 2^{m-n}\langle\langle\Pi_{\hat{b}_{n}\dots \hat{b}_{m-1}}|\mathcal{R}_{B}|\sigma_{\alpha}^{b}\rangle\rangle_{q_{n}}|\rho_{0}\rangle\rangle_{q_{n+1}\dots q_{m-1}}\label{eq:proba_B_frag}
\end{align}

where we have regrouped $\mathcal{R}_{A}\equiv\mathcal{R}_{A}^{\mathrm{a}}\mathcal{R}_{A}^{\mathrm{b}}$
and $\mathcal{R}_{B}\equiv\mathcal{R}_{B}^{\mathrm{a}}\mathcal{R}_{B}^{\mathrm{b}}$.
In other words, $p_{A}^{\alpha}(\hat{b}_{0}\dots \hat{b}_{n-1};b')$ is the probability
of measuring bitstring $\hat{b}_{0}\dots \hat{b}_{n-1},b'$ when measuring the
final state of fragment $A$ with a measurement on axis $\alpha$
for qubit $q_{n}$ (see Fig. \ref{fig:Cutting-sketch}(b)), and $p_{B}^{\alpha b}(\hat{b}_{n}\dots \hat{b}_{m-1})$
is the probability of measuring bitstring $\hat{b}_{n}\dots \hat{b}_{m-1}$ when
measuring the final state of fragment $B$ with qubit $q_{n}$ initially
prepared in the $b$th eigenstate of Pauli matrix $\sigma_{\alpha}$
(see Fig. \ref{fig:Cutting-sketch}(c)).

\paragraph{Variant using Bell pair}

We now derive a different expression based on the following idea:
instead of preparing both eigenstates of $\sigma_{\alpha}$, one can
use an ancilla qubit, prepare a Bell state, and measure the value
of the ancilla along measurement axis $\alpha$ and obtain an equivalent
result, with a slightly different expression.

Switching from the Pauli-basis expression back to the original representation,
Eq. (\ref{eq:proba_B_frag}) is equivalent to
\begin{align*}
p_{B}^{\alpha b}(i) & =\mathrm{Tr}\left[\Pi_{i}\mathcal{O}_{B}(\sigma_{\alpha}^{b}\otimes\rho_{0})\right]
\end{align*}

where $\sigma_{\alpha}^{b}=|\psi_{\alpha}^{b}\rangle\langle\psi_{\alpha}^{b}|$.
Let us decompose
\begin{align*}
|\psi_{\alpha}^{b}\rangle & =\sum_{k\in\{0,1\}}\langle k|\psi_{\alpha}^{b}\rangle|k\rangle
\end{align*}

then
\begin{align*}
  p_{B}^{\alpha b}(i)
 & =\sum_{kk'}\langle k|\psi_{\alpha}^{b}\rangle\langle\psi_{\alpha}^{b}|k'\rangle\mathrm{Tr}\left[\Pi_{i}\cdot\mathcal{O}_{B}(|k\rangle\langle k'|\otimes\rho_{0})\right]\\
 & =\sum_{kk'}\langle\psi_{\alpha}^{b*}|k\rangle\langle k'|\psi_{\alpha}^{b*}\rangle
 \;\;\;\times\mathrm{Tr}\left[\left(I\otimes\Pi_{i}\right)\cdot\left(\mathcal{I}\otimes\mathcal{O}_{B}\right)(I\otimes|k\rangle\langle k'|\otimes\rho_{0})\right]\\
 & =\mathrm{Tr}\Bigg[\left(|\psi_{\alpha}^{b*}\rangle\langle\psi_{\alpha}^{b*}|\otimes\Pi_{i}\right)
 \;\;\;\times\left(\mathcal{I}\otimes\mathcal{O}_{B}\right)\left(\sum_{kk'}|k\rangle\langle k'|\otimes|k\rangle\langle k'|\otimes\rho_{0}\right)\Bigg]\\
 & =2\mathrm{Tr}\left[\left(\Pi_{\alpha}^{b*}\otimes\Pi_{i}\right)\cdot\left(\mathcal{I}\otimes\mathcal{O}_{B}\right)\left(\rho_{\Phi^{+}}\otimes\rho_{0}\right)\right]
\end{align*}

where $\Pi_{\alpha}^{b*}=|\psi_{\alpha}^{b*}\rangle\langle\psi_{\alpha}^{b*}|$
is the projector onto the complex conjugate of the $b$th eigenstate
of the $\sigma_{\alpha}$ Pauli matrix, and $\rho_{\Phi^{+}}$ is
the density matrix of the Bell state
\begin{equation}
|\Phi^{+}\rangle\equiv\frac{1}{\sqrt{2}}\sum_{k=0,1}|kk\rangle.\label{eq:Bell_pair_def}
\end{equation}

In the second line, we have added an ancilla qubit. Now, let us note
that for $\alpha=X,Z$, $|\psi_{\alpha}^{b}\rangle=|\psi_{\alpha}^{b*}\rangle$
(the eigenvector is real-valued), while $|\psi_{Y}^{b*}\rangle=|\psi_{Y}^{1-b}\rangle$,
and let us define 
\begin{equation}
\hat{p}_{B}^{\alpha}(b;i)\equiv\mathrm{Tr}\left[\Pi_{\alpha}^{b}\otimes\Pi_{i}\left(\mathcal{I}\otimes\mathcal{O}_{B}\right)(\rho_{\Phi^{+}}\otimes\rho_{0})\right].\label{eq:proba_b_frag_modified}
\end{equation}

Then:
\begin{align*}
p_{B}^{\alpha b}(i) & =\begin{cases}
2\hat{p}_{B}^{\alpha}(i;b) & \alpha=X,Z\\
2\hat{p}_{B}^{\alpha}(i;1-b) & \alpha=Y
\end{cases}
\end{align*}

Thus, after relabeling $b\rightarrow1-b$ for $\alpha=Y$ in the final
formula Eq. (\ref{eq:final_formula}), we finally obtain the final expression:

\begin{equation}
  \boxed{p(\hat{b}_{0}\dots \hat{b}_{m-1})=\label{eq:final_formula_Bell}
  \sum_{\alpha=X,Y,Z}\sum_{bb'\in\{0,1\}^{2}}\gamma_{\alpha}^{bb'}p_{A}^{\alpha}(\hat{b}_{0}\dots \hat{b}_{n-1};b')p_{B}^{\alpha}(b;\hat{b}_{n}\dots \hat{b}_{m-1}).}
\end{equation}
where $\gamma_{X}^{bb'}=2\delta_{bb'}-1$, $\gamma_{Y}^{bb'}=-\gamma_{X}^{bb'}$
and $\gamma_{Z}^{bb'}=2\delta_{bb'}$.

\begin{figure}
\begin{centering}
\includegraphics[width=0.8\columnwidth]{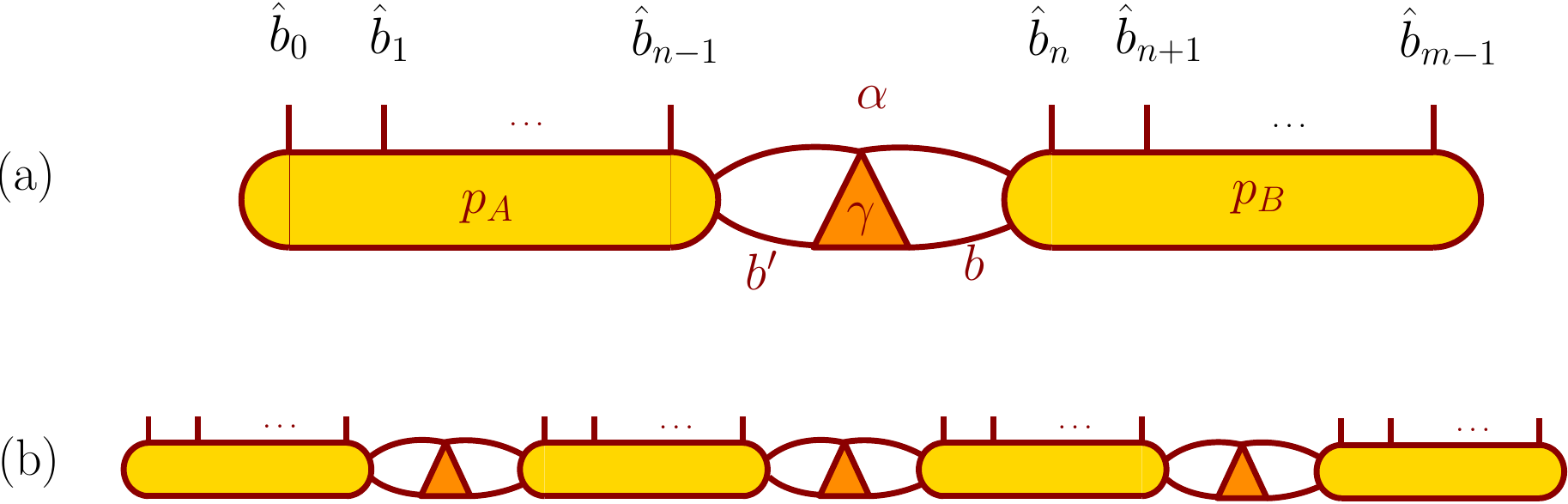}
\par\end{centering}
\caption{Graphical representation of the contraction formula. Panel (a): Two fragment case. Panel (b): Multifragment case for the GHZ circuit shown in Fig.\ref{fig:Cutting-sketch} (a). \label{fig:Contraction_illustration}}
\end{figure}

The graphical representation for such a contraction is shown in Fig.~\ref{fig:Contraction_illustration} (a).

\subsubsection{Multi-fragment case}

\begin{figure}
\begin{centering}
\includegraphics[width=0.8\columnwidth]{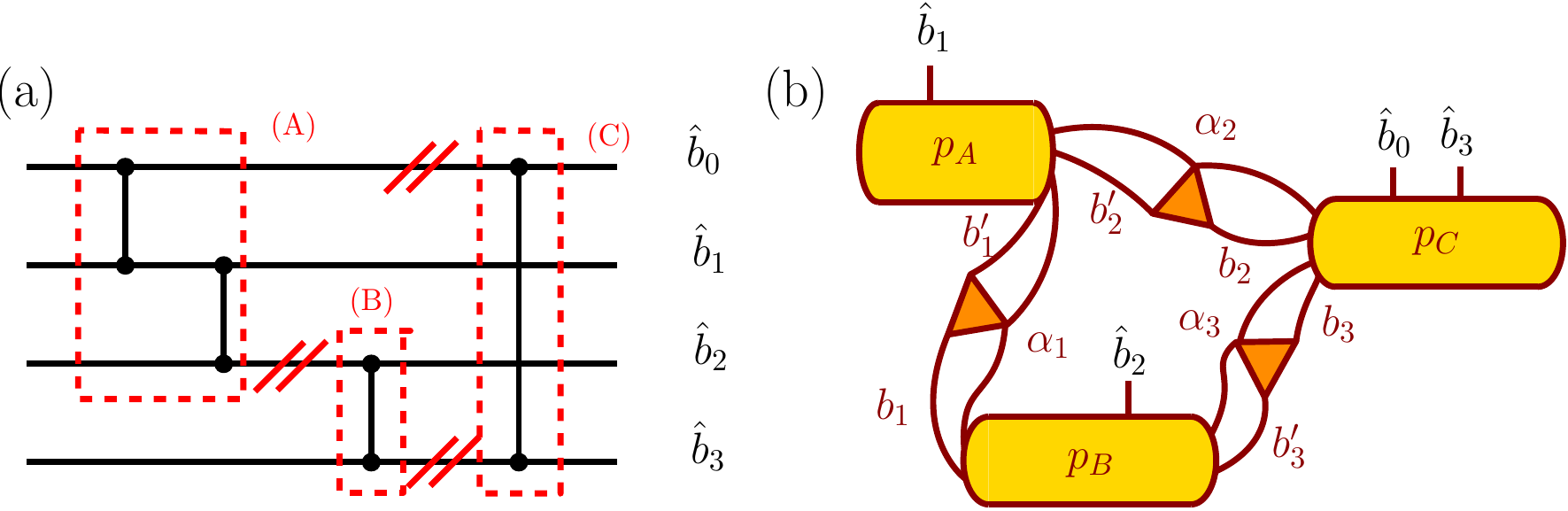}
\par\end{centering}
\caption{Graphical representation of the contraction formula for a generic case (here with three fragments). Panel (a): Sketch of the fragmentation of a four-qubit circuit in three fragments. Panel (b): Corresponding tensor network to contract to get final distribution. \label{fig:Contraction_illustration_three_frags}}
\end{figure}

The formula for the multi-fragment case can be inferred from that
of the two-fragment case: the procedure sketched for the two-fragment
case can be recast in more generic terms, as described in~\cite{Peng2019}.
This is done by considering the directed acyclic graph $G=(V,E)$
corresponding to the quantum circuit at hand (see Fig.~\ref{fig:Contraction_illustration_three_frags} for an illustration of the procedure). Its vertices $V$ are
quantum operations such as qubit initialization, measurement and gates.
The cutting procedure amounts to finding a subset $E'\subset E$ of
$M$ (directed) edges in this graph whose removal leads to $K$ disconnected
directed acyclic graphs $\{G^{(i)}=\left(V_{i},E_{i}\right)\}_{i=1\dots K}$.
In each disconnected graph, $n_{i}+m_{i}$ vertices have a dangling
edge corresponding to the original $n_{i}$ incoming and $m_{i}$
outgoing edges connecting it to the rest of the original graph, with
$\sum_{i}n_{i}=\sum_{i}m_{i}=M$. One then adds a measurement along
axis $\alpha_{k}$ ($\alpha_{k}=X,Y,Z)$ as a termination to each
outgoing dangling edge ($k=1\dots n_{i}$), and a Bell-state gadget
(as described in the previous subsection), whose ancilla line is terminated
by an $\alpha'_{k}$-measurement, to each incoming dangling edge. Translating
the family of graphs $G_{\alpha_{1\dots}\alpha_{n_{i}},\alpha'_{1\dots}\alpha'_{m_{i}}}^{(i)}$back
to quantum circuits $\mathcal{C}_{\alpha_{1\dots}\alpha_{n_{i}},\alpha'_{1\dots}\alpha'_{m_{i}}}^{(i)}$,
we can sample (using a quantum computer) the corresponding probability
distributions. We denote as
\[
p_{i}^{\alpha_{1}\dots\alpha_{n_{i}},\alpha'_{1}\dots\alpha'_{m_{i}}}\left(b_{1},\dots b_{n_{i}};s;b'_{1},\dots b'_{m_{i}}\right)
\]

the probability of measuring bitstring $b_{1},\dots b_{n_{i}};s;b'_{1},\dots b'_{m_{i}}$,
with $s=(\hat{b}_1 \dots \hat{b}_{p_i})$ a bitstring corresponding to the state of ``final'' qubits
of circuit $\mathcal{C}^{(i)}$, and $(b_{1},\dots b_{n_{i}})$ (resp.
$b'_{1},\dots b'_{m_{i}})$) the bitstrings corresponding to the measured
value for the measurements on the incoming (resp. outgoing)
edges of sub-graph $G^{(i)}$ after pre-measurement rotations along
axes $\alpha_{1}\dots\alpha_{n_{i}},\alpha'_{1}\dots\alpha'_{m_{i}}$.

The final probability distribution is obtained by contracting the
tensor network defined by the graph $\hat{G}=\left(\hat{V},\hat{E}\right)$,
with $|\hat{V}|=K+M$ and $|\hat{E}|=2M$. Here, $K$ ``fragment''
vertices correspond to the $K$ disconnected components $\{G^{(i)}\}$,
and $M$ ``connecting'' vertices to the $M$ removed edges. The
$2M$ edges connect each of the $K$ ``fragment'' vertices via one
of the $M$ ``connecting'' vertices. To each ``fragment'' vertex,
we associate a distribution $p_{i}$, while to each ``connecting''
vertex, we associate a $\gamma$ tensor (as defined below Eq. (\ref{eq:final_formula_Bell})).

We give an example of such a tensor network for the Greenberger-Horne-Zeilinger (GHZ) circuit we considered in our previous work as well in Fig.~\ref{fig:Contraction_illustration} (b): in this case, the underlying graph turns out to be linear.
We also show, in Fig.~\ref{fig:Contraction_illustration_three_frags}, an example with a more complex circuit and the resulting, more complex tensor network. Here, $K=3$ and $M=3$.

The contraction of these networks yields the sought-after distribution. The classical complexity of carrying out this contraction will be discussed in section~\ref{subsec:contraction_complexity}.

\subsection{Noisy simulation}
\label{sec:noisy_sim}

NISQ processors are characterized by a substantial level of noise. In this section, we describe what noise processes we took into account in our simulation of the IBM Johannesburg quantum processor.

In this study, we focus on the noise processes whose quantitative levels are reported by the hardware manufacturer, IBM (see Table~\ref{tab:Johannesburg-processor-metrics} for a summary of the numerical values used in the noisy simulations below). This pragmatic approach is justified a posteriori by the reasonable agreement of our numerical simulations with the experimental data (see Ref.~\cite{Ayral2020}, and Section~\ref{results} below). It should nevertheless be emphasized that (i) it uses rather simple noise models, that should be compared to noise models extracted from a full process tomography of the processor, and that (ii) it excludes some noise processes that are suspected to affect the final quantum state distribution in a non-negligible way, e.g., crosstalk (spatially correlated noise) and temporally correlated noise (like 1/f noise). 

\begin{table}
\begin{centering}
\begin{tabular}{|c|c|}
\hline 
Parameter & Value\tabularnewline
\hline 
\hline 
Readout error rate $\gamma$ & 4.1\%\tabularnewline
\hline 
One-qubit-gate error rate $\epsilon_{\mathrm{avg}}^{(1)}$ & 0.041\%\tabularnewline
\hline 
Two-qubit-gate error rate $\epsilon_{\mathrm{avg}}^{(2)}$ & 0.202\%\tabularnewline
\hline 
Relaxation time $T_{1}$ & 65 $\mu$s\tabularnewline
\hline 
Dephasing time $T_{2}$ & 70 $\mu$s\tabularnewline
\hline 
\end{tabular}
\par\end{centering}
\caption{Johannesburg processor metrics, as retrieved from IBM Quantum Experience on March 5th, 2020. All rates are averages over all the qubits/qubit pairs.\label{tab:Johannesburg-processor-metrics}}
\end{table}

The most prominent source of error in today's superconducting processors is the readout error. The duration of the dispersive readout conducted in transmon processors, of the order of a few microseconds, makes for a higher probability of error, most notably of the relaxation (or amplitude damping) type. 
We thus model the readout process as a two-outcome POVM corresponding to an amplitude-damping quantum channel of duration $\tau$ followed by a perfect $Z$-axis measurement: $\lbrace \boldsymbol{I} - \boldsymbol{E}, \boldsymbol{E} \rbrace$, with $\boldsymbol{E}=\left(\begin{array}{cc}
0 & 0\\
0 & 1-\gamma
\end{array}\right).$ The duration $\tau$ is adjusted so as to obtain a readout error rate $\gamma = 1 - e^{-\tau/T_1}$ that matches the readout error rate reported by IBM. With $\gamma = 4.1\%$ and $T_1=65\mu s$, we find $\tau=2.75 \mu s$, a duration that is consistent with the usual measurement durations of dispersive readout processes. 
Note that this noise model does not include measurement crosstalk effects~\cite{Chen2019}.

Another source of error is gate noise, i.e. gate imperfections. Here, since the hardware manufacturer only reports average 1- and 2-qubit gate error rates, we picked the simplest noise process to model gate noise, namely depolarizing noise with a depolarization probability adjusted so that the average process fidelity $\mathcal{F}_\mathrm{avg}$ matches the qubit-averaged average error rates $\epsilon_\mathrm{avg} = 1 - \mathcal{F}_\mathrm{avg}$ reported by the hardware maker.
We recall that the one-qubit depolarizing noise process is characterized by the following Kraus operators:
\begin{align*}
\boldsymbol{K}_{0}^{D} & =\sqrt{1-p_{(1)}^{D}}\boldsymbol{I},\\
\boldsymbol{K}_{i}^{D} & =\sqrt{p_{(1)}^{D}}\boldsymbol{\sigma}_{i},\;\;i=1,2,3
\end{align*}
where $\boldsymbol{\sigma}_{i}$ denote the Pauli spin matrices. We model two-qubit depolarization processes as a tensor product of the one-qubit depolarizing noise.
Let us stress that more structured, and therefore more accurate, noise models could be extracted from quantum process tomography methods, at the cost of a larger characterization overhead. Furthermore, this noise model does not include any crosstalk effects (see e.g~\cite{Sarovar2019}), despite evidence that they play some role in today's NISQ processors.

Finally, we include the effect of decoherence on idle qubits, i.e. qubits that are not being acted upon by a quantum gate, but are nevertheless coupled to the outside environment. This decoherence can be decomposed into two main types, namely relaxation and dephasing. Relaxation (also known as amplitude damping or, in other contexts, spontaneous emission) causes excited qubits (i.e. in state $|1\rangle$) to relax to their ground state ($|0\rangle$) with a probability that is characterized by a time $T_1$: $p_{\tau_{\mathrm{idle}}}^{\mathrm{A.D}}=1-e^{-\tau_{\mathrm{idle}}/T_{1}}$, namely, the longer the idling duration $\tau_{\mathrm{idle}}$, the higher the probability of a relaxation event.
Similarly, dephasing events cause the two components $|0\rangle$ and $|1\rangle$ of a superposed state to acquire an unwanted dephasing with a certain probability. Under simplifying assumptions about the power spectral density (PSD) of the qubit-environment system, namely the assumption of a white noise PSD, this probability is given by  $p_{\tau_{\mathrm{idle}}}^{\mathrm{P.D}}=1-e^{-2\tau_{\mathrm{idle}}/T_{\varphi}}$,
with $\frac{1}{T_{\varphi}}=\frac{1}{T_{2}}-\frac{1}{2T_{1}}$.
We note that this is a quite strong simplification, as actual transmon processors are known to have a PSD that deviates from white noise, with, most notably, a sizable pink (1/f) noise component (see e.g~\cite{Paladino2014} for a review) that leads to a deviation to the exponential decay of the formula we used.
Let us also stress that such a noise modeling does not take into account temporally correlated noise.
As a reminder, here are the Kraus operators associated with amplitude damping and (pure) dephasing:
\begin{align*}
\boldsymbol{K}_{0}^{\mathrm{A.D}} & =\left[\begin{array}{cc}
1 & 0\\
0 & \sqrt{1-p_{\tau_{\mathrm{idle}}}^{\mathrm{A.D}}}
\end{array}\right],\boldsymbol{K}_{1}^{\mathrm{A.D}}=\left[\begin{array}{cc}
0 & \sqrt{p_{\tau_{\mathrm{idle}}}^{\mathrm{A.D}}}\\
0 & 0
\end{array}\right],\\
\boldsymbol{K}_{0}^{\mathrm{P.D}} & =\left[\begin{array}{cc}
1 & 0\\
0 & \sqrt{1-p_{\tau_{\mathrm{idle}}}^{\mathrm{P.D}}}
\end{array}\right],\boldsymbol{K}_{1}^{\mathrm{P.D}}=\left[\begin{array}{cc}
0 & 0\\
0 & \sqrt{p_{\tau_{\mathrm{idle}}}^{\mathrm{P.D}}}
\end{array}\right].
\end{align*}
The values we used for $T_1$ and $T_2$ are reported in Table~\ref{tab:Johannesburg-processor-metrics}.

The noisy simulations are conducted on the Atos Quantum Learning Machine (QLM), a classical supercomputing platform dedicated to writing, simulating and optimizing quantum algorithms \cite{Atos}.

\begin{figure}
\begin{centering}
\includegraphics[width=0.42\columnwidth]{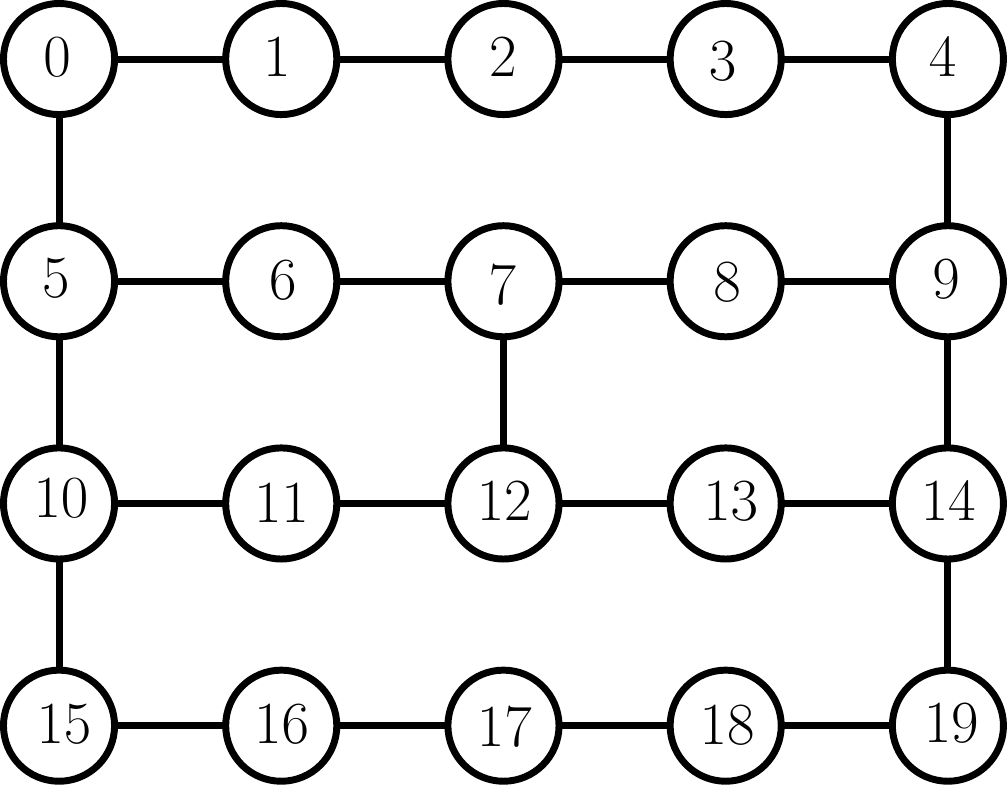}
\par\end{centering}
\caption{Qubit connectivity map of the Johannesburg processor. Edges are shown between qubit pairs coupled via a resonator that allows application of the two-qubit CNOT gate.\label{fig:johannesburg_connectivity}}
\end{figure}

Before simulating the circuits resulting from the fragmentation procedure described in the previous section, we use the QLM's \emph{Nnizer} plugin to compile the circuits, i.e. most notably to adapt them to the Johannesburg processor's restricted qubit topology (shown in Fig.~\ref{fig:johannesburg_connectivity}).
Then, we perform noisy simulations using a density-matrix-based noise simulator that uses a dense representation of the density matrix $\rho$ of the qubit register.

\section{Results}\label{results}

\subsection{Summary of previous results}\label{overview}

\begin{figure}
\begin{centering}
\includegraphics[width=0.75\columnwidth]{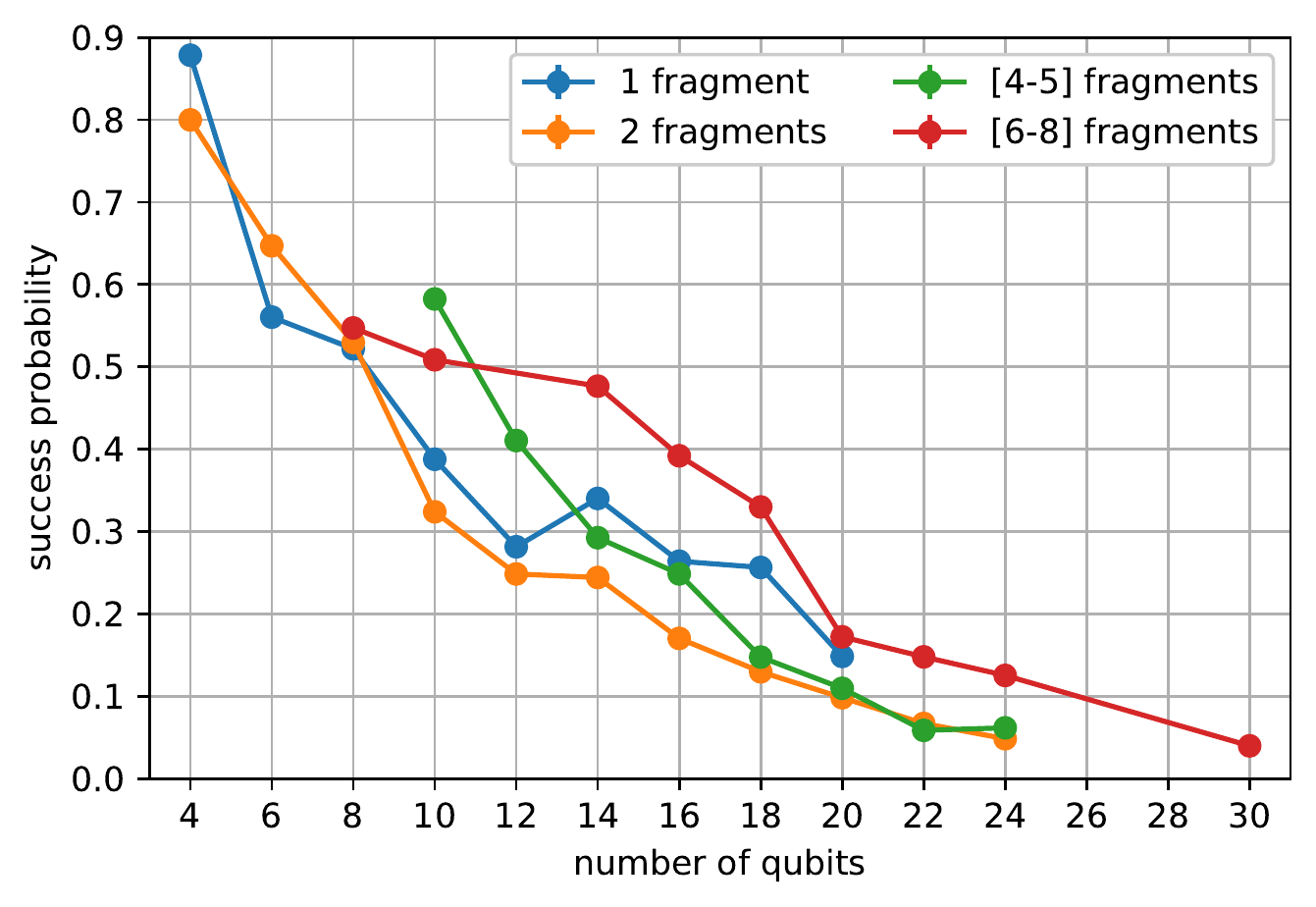}
\par\end{centering}
\caption{Success probability as a function of circuit size (number of qubits)
for various numbers of fragments using IBM's Johannesburg processor.\label{fig:Success-probability-IBM-Johannesburg}}
\end{figure}

\begin{figure}
\begin{centering}
\includegraphics[width=0.9\columnwidth]{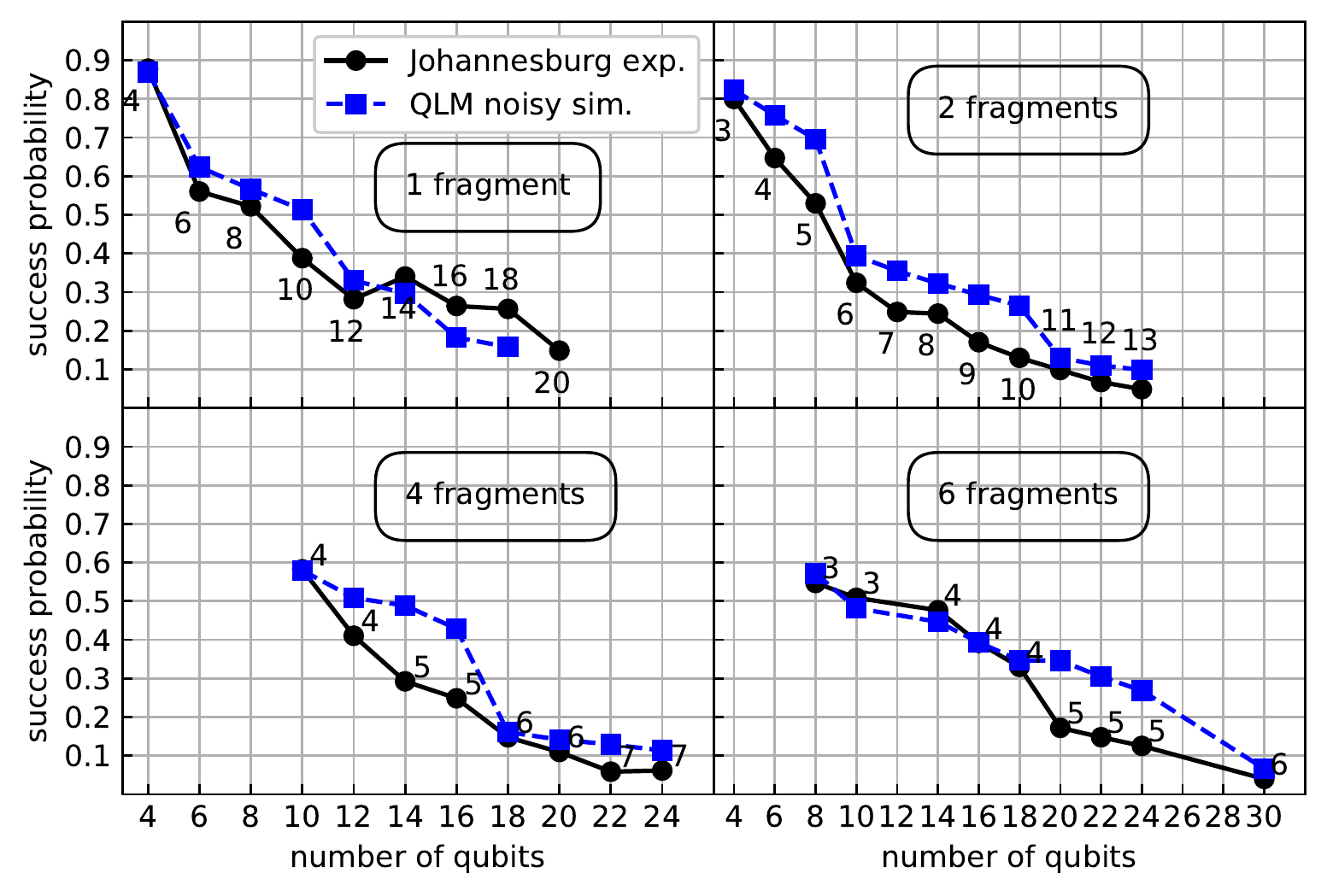}
\par\end{centering}
\caption{Success probability as a function of circuit size (number of qubits)
for various numbers of fragments using IBM's Johannesburg processor
(solid black lines) and Atos QLM noisy simulation (dashed blue lines).
The black integers next to each black disk indicate the maximum fragment
size (in number of qubits).\label{fig:Success-probability-IBM-vs-QLM-Johannesburg}}
\end{figure}

In~\cite{Ayral2020}, we investigated the performance of the circuit-cutting procedure for a simple GHZ-type circuit shown in Fig.~\ref{fig:Cutting-sketch}(a). As a proxy for the quality of the procedure, we chose the quantity
\begin{equation}
P_{\mathrm{success}}\equiv p\left(|0\rangle^{\otimes m/2}|1\rangle^{\otimes m/2}\right)+p\left(|1\rangle^{\otimes m/2}|0\rangle^{\otimes m/2}\right),\label{eq:p_success_def}
\end{equation}
which, given the GHZ circuit at hand, is unity in the absence of any noise.

We carried out the procedure both using an actual 20-qubit processor, IBM Johannesburg, and using the Atos Quantum Learning Machine's noisy simulator.

The experimental success probabilities, shown in Fig.~\ref{fig:Success-probability-IBM-Johannesburg}, display two clear trends: on the one hand, increasing the number of qubits leads to a decreasing success probability. This trend can be accounted for by the fact that increasing the number of qubits increases the number of gates of the circuit, and thus the sensitivity to gate errors and environmental decoherence. On the other hand, increasing the number of fragments in general leads to an improved success probability: the 6-8 fragment success probabilities are larger than the success probabilities obtained for lower numbers of fragments (with some exceptions to this observation: the one-fragment success probability often exceeds that of the 2 and 4-5 fragment cases, maybe due to compiler optimizations on the hardware side for circuits with larger numbers of qubits; we also note a point at $n_\mathrm{qbits}=10$ where the 4-5 fragment success probability exceeds that of the 6-8 fragment case). This trend can be ascribed to the smaller gate count of each individual fragment, and thus a reduced sensitivity to errors. This smaller gate count not only comes from the mere cutting procedure, but also from the fact that smaller circuits better match the limited connectivity (Fig.~\ref{fig:johannesburg_connectivity}) of the Johannesburg chip. Conversely, larger circuits need to be compiled to fulfill the connectivity constraints, leading to larger gate counts.

To substantiate these interpretations, we performed noisy simulations with noise models constructed using the constructor's calibration data (Table~\ref{tab:Johannesburg-processor-metrics}). We show the results in Fig.~\ref{fig:Success-probability-IBM-vs-QLM-Johannesburg}: a 20\% agreement is found between the noisy simulations and the experimental data.
In particular, the drops in success probability, which can be traced back to connectivity-related insertions of SWAP gates, are reproduced.
We note that the error bars coming from the finite number of shots (8192) used for each fragment are contained within the data symbols.

\subsection{Analysis of the influence of the different noise types}

\begin{figure}
\begin{centering}
\includegraphics[width=0.9\columnwidth]{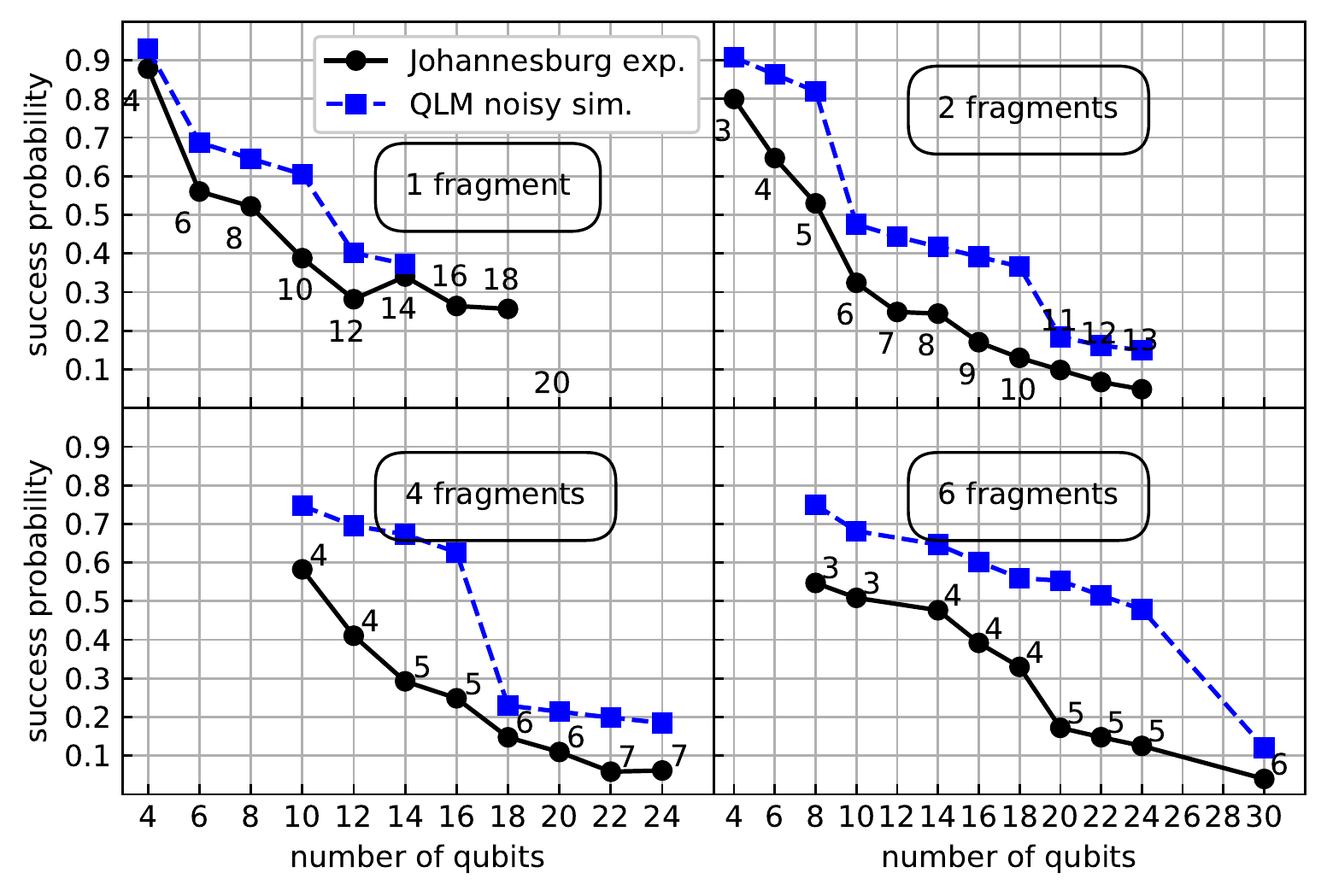}
\par\end{centering}
\caption{
\emph{Effect of better readout:}
Same as Fig.~\ref{fig:Success-probability-IBM-Johannesburg}, but with a readout duration divided by 5.\label{fig:Success-probability-IBM-vs-QLM-Johannesburg-faster-readout}}
\end{figure}

\begin{figure}
\begin{centering}
\includegraphics[width=.9\columnwidth]{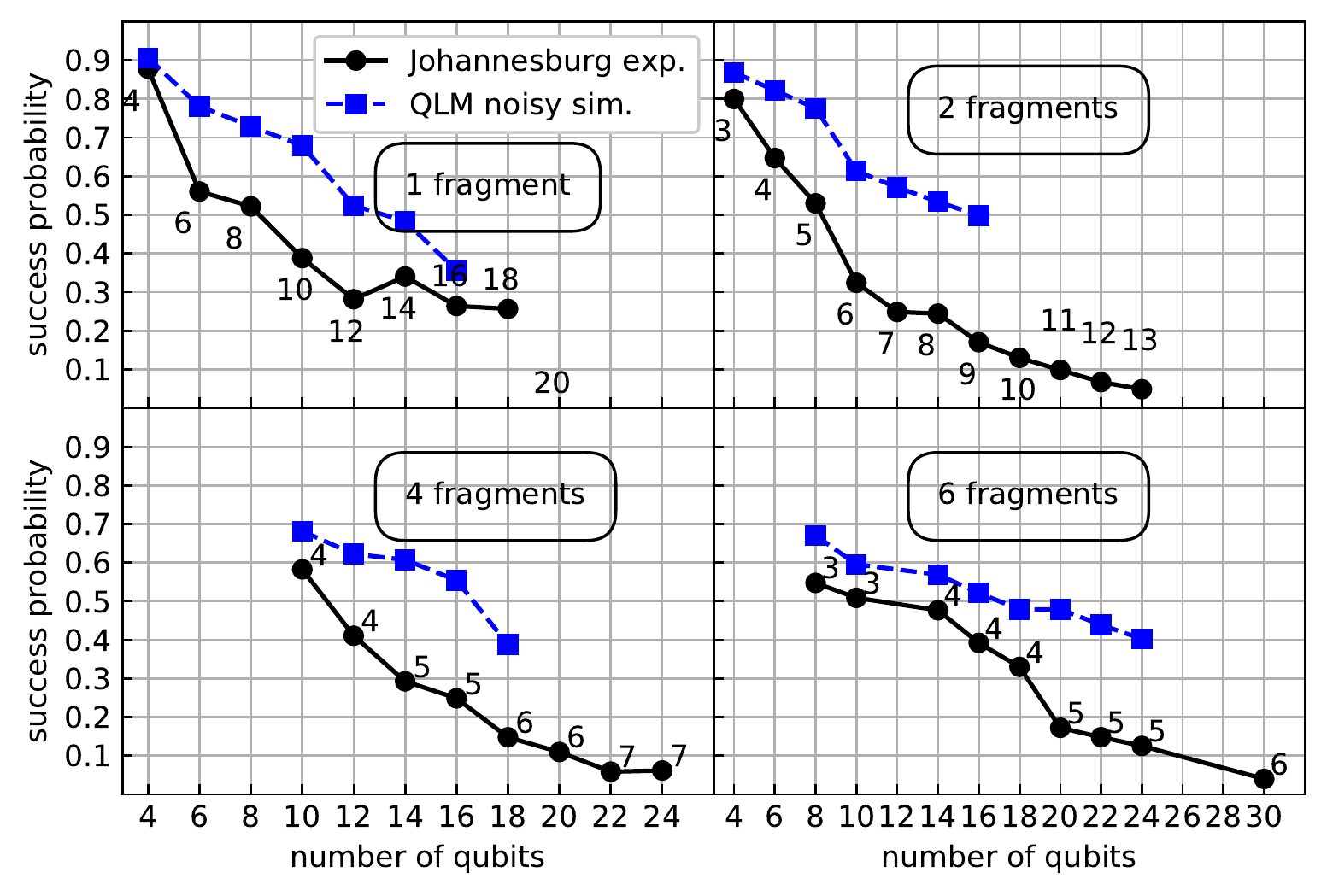}
\par\end{centering}
\caption{
\emph{Effect of better gates:}
Same as Fig.~\ref{fig:Success-probability-IBM-Johannesburg}, but with a depolarizing error per gate divided by 5.\label{fig:Success-probability-IBM-vs-QLM-Johannesburg-better-gates}}
\end{figure}

\begin{figure}
\begin{centering}
\includegraphics[width=.9\columnwidth]{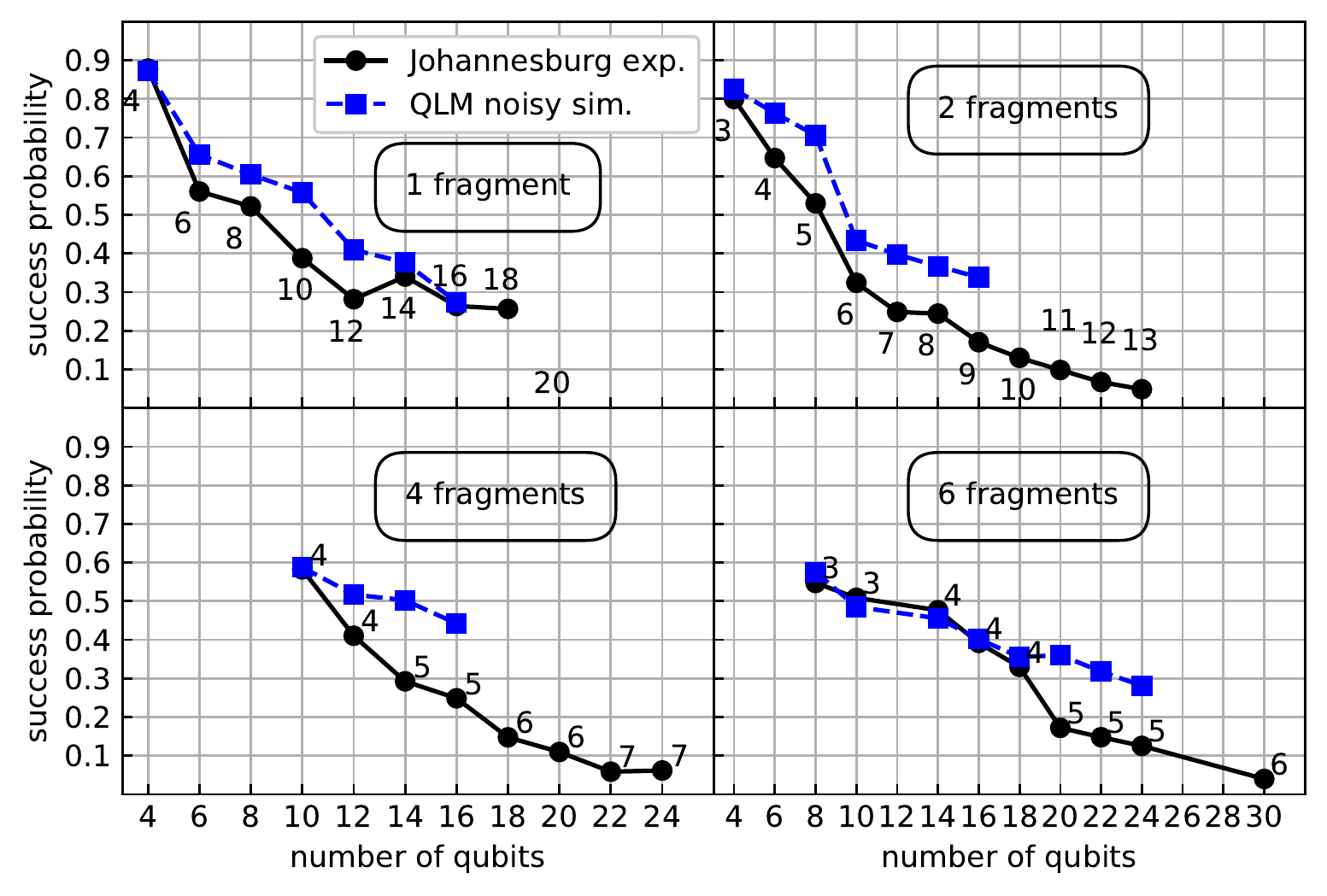}
\par\end{centering}
\caption{
\emph{Effect of better coherence:}
Same as Fig.~\ref{fig:Success-probability-IBM-Johannesburg}, but with  $T_1$ and $T_2$ coherence times multiplied by 5.\label{fig:Success-probability-IBM-vs-QLM-Johannesburg-better-coherence}}
\end{figure}

\begin{figure}
\begin{centering}
\includegraphics[width=0.75\columnwidth]{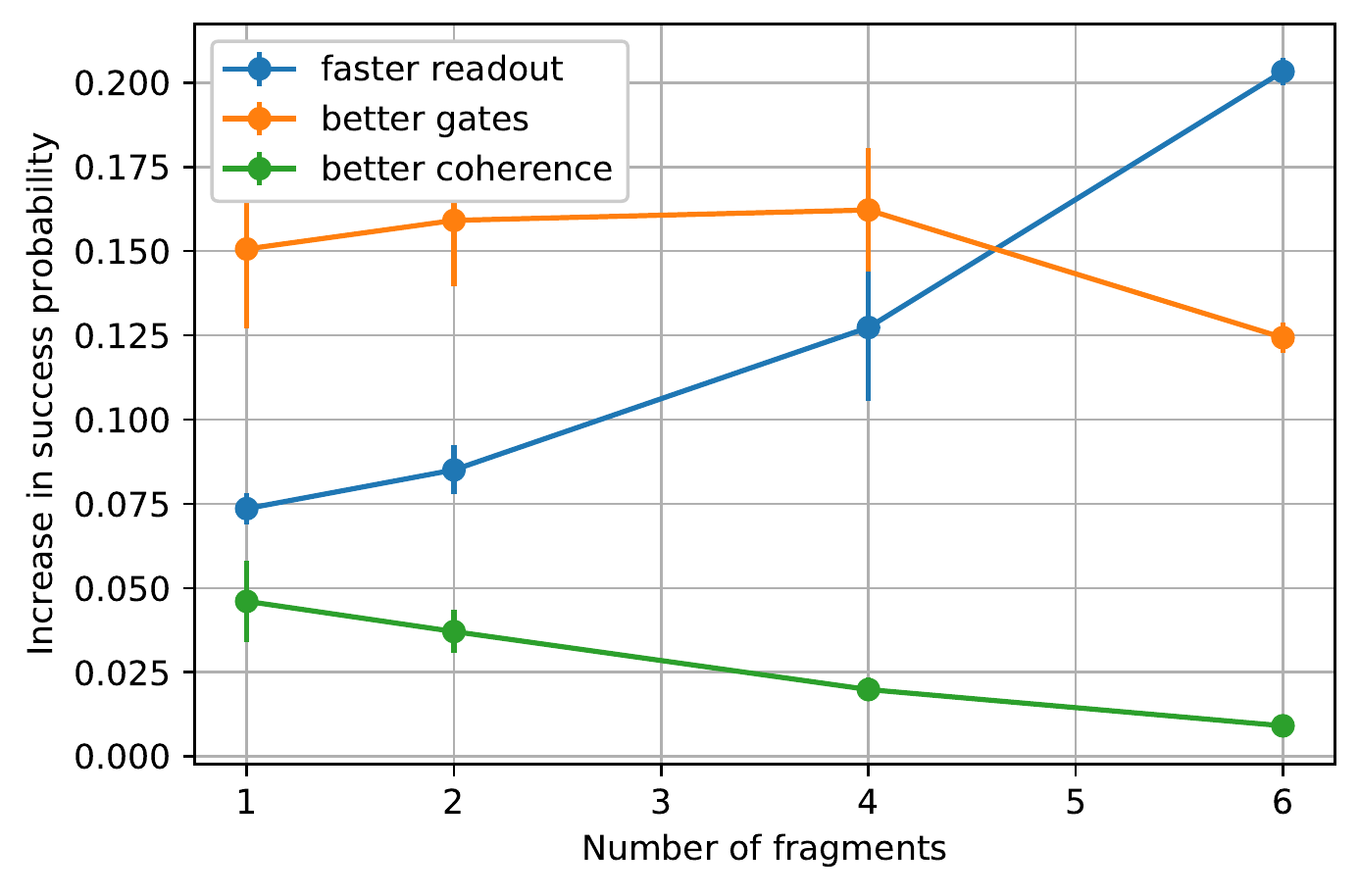}
\par\end{centering}
\caption{Increase in success probability averaged over the number of qubits as a function of the number of fragments, $\Delta P (\mathcal{S}, n_\mathrm{f})$,
for a faster readout (blue, parameters of Fig.~\ref{fig:Success-probability-IBM-vs-QLM-Johannesburg-faster-readout}),
better gates (orange, parameters of Fig.~\ref{fig:Success-probability-IBM-vs-QLM-Johannesburg-better-gates}),
better coherence (green, parameters of Fig.~\ref{fig:Success-probability-IBM-vs-QLM-Johannesburg-better-coherence}).
\label{fig:Success-probability-diff-various-scenarios}}
\end{figure}

In this section, we study and compare the differential impact of all the noise types we have previously taken into account: gate imperfections, idling and readout errors. 
Our goal is to understand which types of noise have a particularly severe influence on the fidelity of the fragmenting procedure and to formulate recommendations as to which noise types should be addressed first if one wants to make the most of the fragmenting procedure.
Hence, we study the influence of the three noise types by simulating better readout measurements (Fig. \ref{fig:Success-probability-IBM-vs-QLM-Johannesburg-faster-readout}), better gates (Fig. \ref{fig:Success-probability-IBM-vs-QLM-Johannesburg-better-gates}) and a better coherence time (Fig. \ref{fig:Success-probability-IBM-vs-QLM-Johannesburg-better-coherence}).

\paragraph{Faster readout.}
First, we analyze the impact of readout errors by decreasing the duration $\tau$ of the measurements on all the subcircuits generated by the splitting procedure. 
Readout error is at present the largest source of errors in superconducting processors, with error rates as high as a few percent. It is thus reasonable to assume that large experimental efforts are going to be made to reduce this error rate. Here, we suppose the reduction in readout error rate to originate from a reduction of the readout duration (in practice by a factor 5), although it would be equivalent, in this noise model that assumes the errors to come only from an amplitude damping noise, to keep the readout duration fixed and to increase the T1 coherence time (by the same  factor 5). In reality, progress is being made on both fronts (see e.g \cite[Fig.~2.c]{Kjaergaard2020}, for the increasing T1 trend, and \cite{Heinsoo2018} for recent efforts towards faster measurements).

We see in Fig.~\ref{fig:Success-probability-IBM-vs-QLM-Johannesburg-better-coherence} that better readout improves the overall success probability all the more as the fragment number is large.
The difference between the solid and the dashed lines qualitatively increases with the number of readout measurements used, and consequently the number of fragments.
Indeed, more fragments necessitate more measurements to characterize the quantum state of each fragment.
Nevertheless, we still see drops in the evolution of the success probability with the number of qubits. 
It can be explained by the topology constraints that require the use of several SWAP gates when we try to perform gates between physical qubits that are not adjacent.
This calls the study of the next paragraph.

\paragraph{Better gates.}
To model the use of better gates, we choose to lower the amplitude of the depolarizing channel by dividing the depolarizing error rate by a factor of 5. The limited gate fidelity is the second major source of errors in superconducting processors. It comes from calibration errors as well as decoherence. Here, we mimic the improvement in gate quality by simply dividing the error rate by a factor of 5. Such a factor is realistic, in view of the improvements in gate qualities of superconducting processors in the recent years, and of the variability in the error rates reported by the hardware providers (the two-qubit error rates reported for IBM Johannesburg \cite{IBMQX}, Google Sycamore \cite[Fig.2, Table II]{Arute2019} and Rigetti Aspen 7 \cite{RigettiWebSite}, are, respectively, 0.2\%, 0.62\% and 4.8\%).

The results of this change in the noise model can be seen in Fig. ~\ref{fig:Success-probability-IBM-vs-QLM-Johannesburg}. 
We notice that the slope is more regular as the number of qubits increases.
Indeed, a smoothing of the "drops" in success probability is observed.
These drops were the consequence of performing a gate between qubits that are not adjacent in the connectivity map (Fig.~\ref{fig:johannesburg_connectivity}) and that require using several SWAP gates.
Thus, better gates help mitigate the effect of topology.
The insertion of additional SWAP gates because of topology constraints becomes less detrimental to the overall success probability when the inserted gates are of good fidelity. 

\paragraph{Better coherence.}
Finally, in order to understand the impact of coherence on the splitting procedure, we increase the relaxation time $T_1$ and the dephasing time $T_2$ by multiplying them by a factor of 5 (see Sec. \ref{sec:noisy_sim} for a definition of the corresponding Kraus operators).
Decoherence errors indeed account for another portion of the errors incurred by a quantum processor. They not only lead to a decrease in gate fidelity, but also affect idle qubits. Here, the factor of 5 we chose is compatible with the improvements of the recent years (see \cite{Kjaergaard2020}, Fig 2c for the increasing T1 trend)
Doing this will delay both spontaneous emission (amplitude damping) and phase flip (dephasing) events.

As shown in Fig.~\ref{fig:Success-probability-IBM-vs-QLM-Johannesburg-better-coherence}, better coherence only has a limited impact on the fragmenting procedure: 
it seems to improve more the success probability of the runs with fewer fragments than the one of the runs with more fragments where the solid and dashed lines are closer one to the other.
This behavior is expected. 
Using a larger number of fragments imply that the fragments are smaller in terms of qubits size and such small fragments are less sensitive to decoherence.

All these observations are summarized in Fig.~\ref{fig:Success-probability-diff-various-scenarios}, which shows the increased success probability using the new parameters compared to the success probabilities $P_\mathrm{success}^{(0)}$ computed with the Johannesburg noise parameters.
For each of the scenarios $\mathcal{S}$ introduced above, we compute the increase in probability defined as:

\begin{equation}
    \Delta P (\mathcal{S}, n_\mathrm{f}) = \langle P_\mathrm{success}(\mathcal{S}, n_\mathrm{f}, n_\mathrm{q}) - P_\mathrm{success}^{(0)}(n_\mathrm{f}, n_\mathrm{q})\rangle_{n_\mathrm{q}}.
\end{equation}

We see that, as discussed above, better readout is all the more helpful as the number of fragments is large, while, conversely, better coherence is more beneficial for smaller number of fragments.
Achieving better gate fidelities, on the other hand, is equally beneficial with and without fragmentation since the slope of the orange line is close to $1$.
(We stress that because of the arbitrariness in the quantitative choice of level of improvement for the three scenarios, one cannot conclude any quantitative insight from the value of the improvement; here, our conclusions are qualitative and only based on the slope with respect to the number of fragments).
Consequently, to make the most of the fragmenting procedure in the case of numerous fragments, the major error source to focus on is the measurement error by designing faster readouts.

\subsection{Contraction complexity and relation to tensor-network simulation}\label{subsec:contraction_complexity}

\begin{figure}
\begin{centering}
\includegraphics[width=0.85\columnwidth]{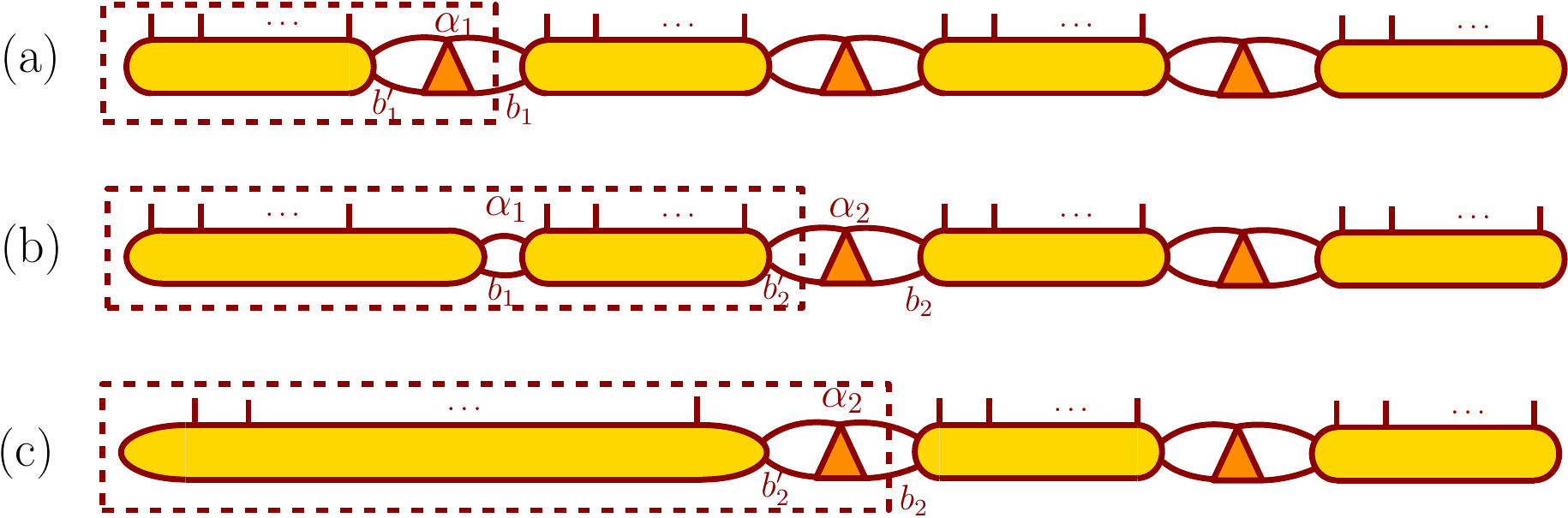}
\par\end{centering}
\caption{First three contraction steps for the fragmentation of the GHZ-type circuit of Fig.~\ref{fig:Cutting-sketch}.
\label{fig:contraction_steps}}
\end{figure}

In this section, we elaborate on the complexity of the fragmentation algorithm. 
As presented in Section~\ref{subsec:Circuit-cutting}, the fragmentation method consists of a quantum and a classical step.
In the quantum step, a batch of quantum circuits is executed on a Quantum Processing Unit (QPU). The number of such circuits scales as the number $K$ of disconnected subgraphs of the original directed acyclic graph with some edges removed. The outcome of this step is a list of probability distributions $p_i$.
In the classical step, a tensor network with nodes corresponding either to the probability distributions or to the $\gamma$ tensors defined in Section~\ref{subsec:Circuit-cutting} needs to be contracted.

Here, we shall be interested in the contraction complexity of such a tensor network, assuming one wants to recover the probability of a single bitstring $(\hat{b}_0, \dots \hat{b}_{m-1})$, i.e. for a fixed assignment of the external legs of the tensor network shown in Fig.~\ref{fig:Contraction_illustration} (b). 
A naive contraction of the tensor network at hand, namely a simultaneous summation over all internal indices ${(\alpha_i, b_i, b'_i)}_{i=1\dots K-1}$, would entail a contraction complexity of $12^{K-1}$, i.e. a classical computation that is exponential in the number $K$ of fragments.
In our case, however, the linear structure of the graph underlying the tensor network allows for a much more efficient sequential contraction strategy.
Such a strategy, which is also widely exploited for contracting so-called Matrix Product States (see e.g.~\cite{Schollwoeck2011,Orus2014} for a review), consists in sequentially contracting the nodes of the network starting from one end of the linear graph.
This is illustrated in Fig.~\ref{fig:contraction_steps}, where we show the first three steps. The contraction complexity of the successive steps is {12, 36, 12, 36, \dots, 12, 36, 12, 6}. For $K$ fragments, this yields an overall contraction complexity of $48 (K-2) + 18 = 48 K - 78$, i.e a linear complexity in the fragment number $K$.

In the case of a general tensor network, the optimal contraction complexity can be shown to be at least of the order of $O(e^T)$, where $T$ is the so-called treewidth of the network graph~\cite{markov2008simulating}. The treewidth of a graph can be defined as a combinatorial metric of closeness of the graph to a tree. There are a few ways to define the treewidth in more formal way: the minimum $k$ for which a given graph is a partial $k$-tree, or the elimination width.

Tensor-network theory can also be leveraged to simulate quantum circuits classically. There are a number of tensor-network-based simulators developed for such simulations: QFlex~\cite{villalonga2019flexible}, AC-QDP~\cite{huang2019alibaba}, Quimb~\cite{gray2018quimb}, and QTensor~\cite{QTensor}. These simulators are typically much faster and more efficient than state vector simulators for shallow circuits~\cite{wu2019full} such as the circuits in this work.
In these tensor simulators, the circuits are not directly represented by tensors, but rather use line graphs, which was proposed by Boixo et al.~\cite{boixo2017simulation}. This approach has multiple benefits. The only disadvantage of the line graph approach is that it has limited usability to simulate sub-tensors of amplitudes, which was resolved in the work by Schutski et al.~\cite{schutski2019adaptive}.

The method studied in our work, circuit cutting, has a counterpart in tensor-network-based simulation.
It is called \emph{tensor slicing}. One way to understand the slice of a tensor as an index that can be viewed as 
the function of many variables evaluated at some value of one variable:
$$f(x_1, x_2, \dots x_n)|_{x_1 = a} = \tilde f(x_2,\dots x_n),$$ 
where variables can have integer values $x_i \in [0,d-1]$.
Thus, in this technique, slicing reduces the number of indices of the tensor one by one. Since all sizes of indices we use are equal to 2,
removal of $n$ vertices allows to split the expression into $2^n$ separate parts. This operation is also equivalent to decomposition of the full tensor expression. 
Each separate tensor is represented by a graph with lower connectivity than
the original one. 
As a result, it dramatically reduces the complexity of finding the optimal elimination. 
Thus, it results in a lower contraction cost. 
It is a powerful technique that allows to simulate large circuits as does the circuit-cutting technique described in this work. 

\subsection{Homogeneous and heterogeneous quantum computing}

One exciting application of the circuit-cutting technique is to allow to execute much larger circuits. It can be done in two ways: split circuits and run sequentially on a quantum device (as we demonstrated in \cite{Ayral2020}), or run at the same time on multiple quantum devices.
The latter way can lead to an exciting new era of how quantum computation is done - distributed quantum computing. It can potentially not only allow for the execution of larger circuits, but also for a much faster execution. It is arguably a more realistic approach in the near future compared to the "true" distributed quantum computing that requires a quantum network connecting quantum devices.
In our approach, indeed, we would utilize only the classical network.

\section{Conclusions}\label{conclusions}

In this work, we further investigated the Quantum Divide and Conquer approach whose first implementation was demonstrated in a recent work of ours \cite{Ayral2020}.

After giving more details as to the mathematical framework and physical models used for this implementation, 
we analyzed the influence of different noise sources on the success probability of a simple, GHZ-type circuit using classical noisy simulations on the Atos Quantum Learning Machine.
We focused on the three main noise sources of today's superconducting processors, namely readout errors, gate errors and decoherence (relaxation and dephasing) on idle qubits.
We showed that readout errors are the most detrimental to the QDC procedure, because QDC requires additional measurements as the number of fragments increases.
Conversely, the effect of idling noise is mitigated by QDC, as QDC results in smaller circuits that are less susceptible to this source of noise.

We also analyzed the computational complexity of QDC using tensor-network methods. 
While for a general circuit the contraction complexity increases exponentially with the number of cuts, for the GHZ-like circuit we studied, the complexity increases linearly with the number of cuts.

Finding more complex circuits in which the contraction complexity is still manageable is an interesting future direction.
Circuits that have a "clustered" structure~\cite{perlin2020quantum}, that are e.g required in methods like the Dynamic Quantum Variational Ansatz~\cite{saleem2020approaches}, are promising candidates. In these methods, indeed, the ansatz has a mixer unitary that is made up of partial mixers that can have limited connectivity between each other, and can therefore form clusters.

\begin{acknowledgements}

This research used resources of the Oak Ridge Leadership Computing
Facility, which is a DOE Office of Science User Facility supported
under Contract DE-AC05-00OR22725. This research also used the resources
of the Argonne Leadership Computing Facility, which is DOE Office
of Science User Facility supported under Contract DE-AC02-06CH11357.
Yuri Alexeev, Zain H. Saleem and Martin Suchara were supported by
the DOE, Office of Science, under Contract DE-AC02-06CH11357. The
compilation and noisy simulations were performed using Argonne National
Laboratory's and Atos Quantum Laboratory's Quantum Learning Machines.

The submitted manuscript has been created by UChicago Argonne, LLC,
Operator of Argonne National Laboratory (\textquotedbl Argonne\textquotedblright ).
Argonne, a U.S. Department of Energy Office of Science laboratory,
is operated under Contract No. DE-AC02-06CH11357. The U.S. Government
retains for itself, and others acting on its behalf, a paid-up nonexclusive,
irrevocable worldwide license in said article to reproduce, prepare
derivative works, distribute copies to the public, and perform publicly
and display publicly, by or on behalf of the Government. The Department
of Energy will provide public access to these results of federally
sponsored research in accordance with the DOE Public Access Plan (http://energy.gov/downloads/doe-public-access-plan).

\textbf{Conflict of Interest: }
The authors declare that they have no conflict of interest.

The data and materials presented in this paper are available upon request to the authors.

\end{acknowledgements}

\appendix

\section{Pauli-basis representation of operators and superoperators}\label{sec:Pauli-basis-representation}

We can decompose any Hermitian operator (including density matrices)
as
\begin{equation}
\rho  =\sum_{\alpha}\rho_{\alpha}P_{\alpha}, \;\;\;\; 
\rho_{\alpha} =\frac{1}{d}\mathrm{Tr}\left[P_{\alpha}\rho\right]
\end{equation}

with $d=2^{n_{\mathrm{qbits}}}$ and $P_{\alpha}$ a generalized Pauli
matrix on $n_{\mathrm{qbits}}$ qubits. Similary, superoperators can
be decomposed on this basis,
\[
\left[\mathcal{R}\right]_{\alpha\beta}=\frac{1}{d}\mathrm{Tr}\left[P_{\alpha}\cdot\mathcal{O}(P_{\beta})\right]
\]

$\mathcal{R}$ is called the Pauli transfer matrix (PTM) representation
of $\mathcal{O}$. Then, the coordinates of $\rho'=\mathcal{O}(\rho)$
is the Pauli basis are simply given by
\begin{equation}
\rho_{\alpha}'  =\frac{1}{d}\mathrm{Tr}\left[P_{\alpha}\mathcal{O}(\rho)\right]
 =\frac{1}{d}\sum_{\beta}\rho_{\beta}\mathrm{Tr}\left[P_{\alpha}\mathcal{O}(P_{\beta})\right]
  =\sum_{\beta}\mathcal{R}_{\alpha\beta}\rho_{\beta}
\end{equation}

We note that
\begin{equation}
\mathrm{Tr}\left[A^{\dagger}\cdot B\right]  =\sum_{\alpha\beta}A_{\alpha}^{*}B_{\beta}\mathrm{Tr}\left[P_{\alpha}^{\dagger}P_{\beta}\right]
  =\sum_{\alpha\beta}A_{\alpha}^{*}B_{\beta}d\delta_{\alpha\beta}
  =d\sum_{\alpha}A_{\alpha}^{*}B_{\alpha}
\end{equation}

Defining the scalar product $\langle\langle A|B\rangle\rangle\equiv\sum_{\alpha}A_{\alpha}^{*}B_{\alpha}$,
we thus have
\begin{equation}
\mathrm{Tr}\left[A^{\dagger}\cdot B\right]=d\langle\langle A|B\rangle\rangle=2^{n_{\mathrm{qbits}}}\langle\langle A|B\rangle\rangle\label{eq:trace_scalar_relation}
\end{equation}

\bibliographystyle{IEEEtran}
\bibliography{fragmenting_long}

\begin{thebibliography}{10}
\providecommand{\url}[1]{#1}
\csname url@samestyle\endcsname
\providecommand{\newblock}{\relax}
\providecommand{\bibinfo}[2]{#2}
\providecommand{\BIBentrySTDinterwordspacing}{\spaceskip=0pt\relax}
\providecommand{\BIBentryALTinterwordstretchfactor}{4}
\providecommand{\BIBentryALTinterwordspacing}{\spaceskip=\fontdimen2\font plus
\BIBentryALTinterwordstretchfactor\fontdimen3\font minus
  \fontdimen4\font\relax}
\providecommand{\BIBforeignlanguage}[2]{{%
\expandafter\ifx\csname l@#1\endcsname\relax
\typeout{** WARNING: IEEEtran.bst: No hyphenation pattern has been}%
\typeout{** loaded for the language `#1'. Using the pattern for}%
\typeout{** the default language instead.}%
\else
\language=\csname l@#1\endcsname
\fi
#2}}
\providecommand{\BIBdecl}{\relax}
\BIBdecl

\bibitem{Ayral2020}
\BIBentryALTinterwordspacing
T.~Ayral, F.~M. {Le Regent}, Z.~Saleem, Y.~Alexeev, and M.~Suchara, ``{Quantum
  divide and compute: Hardware demonstrations and noisy simulations},''
  \emph{Proceedings of IEEE Computer Society Annual Symposium on VLSI, ISVLSI},
  pp. 138--140, 2020. [Online]. Available:
  \url{https://doi.org/10.1109/ISVLSI49217.2020.00034}
\BIBentrySTDinterwordspacing

\bibitem{preskill2018}
\BIBentryALTinterwordspacing
J.~Preskill, ``{Quantum Computing in the NISQ era and beyond},''
  \emph{Quantum}, vol.~2, p.~79, Aug. 2018. [Online]. Available:
  \url{http://dx.doi.org/10.22331/q-2018-08-06-79}
\BIBentrySTDinterwordspacing

\bibitem{Bristlecone_72}
J.~Kelly, ``A preview of {B}ristlecone, {G}oogle's new quantum processor,''
  {Google} {AI} {B}log,
  \url{https://ai.googleblog.com/2018/03/a-preview-of-bristlecone-googles-new.html},
  Mar 2018.

\bibitem{IBM_50}
W.~Knight, ``{IBM} raises the bar with a 50-qubit quantum computer,'' {MIT}
  {T}echnology {R}eview,
  \url{https://www.technologyreview.com/s/609451/ibm-raises-the-bar-with-a-50-qubit-quantum-computer/},
  Nov 2017.

\bibitem{Intel_49}
J.~Hsu, ``{CES} 2018: {I}ntel's 49-qubit chip shoots for quantum supermacy,''
  {IEEE} {S}pectrum,
  \url{https://spectrum.ieee.org/tech-talk/computing/hardware/intels-49qubit-chip-aims-for-quantum-supremacy},
  Jan 2018.

\bibitem{scaling_superconducting}
\BIBentryALTinterwordspacing
J.~M. Gambetta, J.~M. Chow, and M.~Steffen, ``Building logical qubits in a
  superconducting quantum computing system,'' \emph{npj Quantum Information},
  vol.~3, no.~1, p.~2, 2017. [Online]. Available:
  \url{https://doi.org/10.1038/s41534-016-0004-0}
\BIBentrySTDinterwordspacing

\bibitem{Monroe1164}
\BIBentryALTinterwordspacing
C.~Monroe and J.~Kim, ``Scaling the ion trap quantum processor,''
  \emph{Science}, vol. 339, no. 6124, pp. 1164--1169, 2013. [Online].
  Available: \url{https://science.sciencemag.org/content/339/6124/1164}
\BIBentrySTDinterwordspacing

\bibitem{Saffman_neutral_atoms}
\BIBentryALTinterwordspacing
M.~Saffman, ``{Quantum computing with neutral atoms},'' \emph{National Science
  Review}, vol.~6, no.~1, pp. 24--25, Sep. 2018. [Online]. Available:
  \url{https://doi.org/10.1093/nsr/nwy088}
\BIBentrySTDinterwordspacing

\bibitem{entanglement_defect_spins}
\BIBentryALTinterwordspacing
F.~Dolde, I.~Jakobi, B.~Naydenov, N.~Zhao, S.~Pezzagna, C.~Trautmann,
  J.~Meijer, P.~Neumann, F.~Jelezko, and J.~Wrachtrup, ``Room-temperature
  entanglement between single defect spins in diamond,'' \emph{Nature Physics},
  vol.~9, pp. 139 EP --, Feb. 2013. [Online]. Available:
  \url{https://doi.org/10.1038/nphys2545}
\BIBentrySTDinterwordspacing

\bibitem{Heralded_entanglement}
\BIBentryALTinterwordspacing
H.~Bernien, B.~Hensen, W.~Pfaff, G.~Koolstra, M.~S. Blok, L.~Robledo, T.~H.
  Taminiau, M.~Markham, D.~J. Twitchen, L.~Childress, and R.~Hanson, ``Heralded
  entanglement between solid-state qubits separated by three metres,''
  \emph{Nature}, vol. 497, pp. 86 EP --, Apr. 2013. [Online]. Available:
  \url{https://doi.org/10.1038/nature12016}
\BIBentrySTDinterwordspacing

\bibitem{Bravyi2016}
\BIBentryALTinterwordspacing
S.~Bravyi, G.~Smith, and J.~A. Smolin, ``Trading classical and quantum
  computational resources,'' \emph{Phys. Rev. X}, vol.~6, p. 021043, Jun. 2016.
  [Online]. Available: \url{https://link.aps.org/doi/10.1103/PhysRevX.6.021043}
\BIBentrySTDinterwordspacing

\bibitem{Peng2019}
\BIBentryALTinterwordspacing
T.~Peng, A.~Harrow, M.~Ozols, and X.~Wu, ``Simulating large quantum circuits on
  a small quantum computer,'' \emph{arXiv preprint arXiv:1904.00102}, 2019.
  [Online]. Available: \url{https://arxiv.org/abs/1904.00102}
\BIBentrySTDinterwordspacing

\bibitem{Cross2019}
\BIBentryALTinterwordspacing
A.~W. Cross, L.~S. Bishop, S.~Sheldon, P.~D. Nation, and J.~M. Gambetta,
  ``{Validating quantum computers using randomized model circuits},''
  \emph{Physical Review A}, vol. 100, no.~3, p. 032328, Sep. 2019. [Online].
  Available: \url{http://dx.doi.org/10.1103/PhysRevA.100.032328}
\BIBentrySTDinterwordspacing

\bibitem{perlin2020quantum}
\BIBentryALTinterwordspacing
M.~A. Perlin, Z.~H. Saleem, M.~Suchara, and J.~C. Osborn, ``Quantum circuits:
  Divide and compute with maximum likelihood tomography,'' \emph{arXiv preprint
  arXiv:2005.12702}, 2020. [Online]. Available:
  \url{https://arxiv.org/abs/2005.12702}
\BIBentrySTDinterwordspacing

\bibitem{Tang_CutQC}
W.~Tang, T.~Tomesh, J.~Larson, M.~Suchara, and M.~Martonosi, ``{CutQC:} using
  small quantum computers for large quantum circuit evaluations,'' in
  \emph{Proceedings of the ACM International Conference on Architectural
  Support for Programming Languages and Operating Systems (ASPLOS)}, 2021.

\bibitem{1203.5813}
\BIBentryALTinterwordspacing
J.~Preskill, ``Quantum computing and the entanglement frontier,''
  \emph{arXiv:1203.5813}, Nov 2012. [Online]. Available:
  \url{https://arxiv.org/abs/1203.5813}
\BIBentrySTDinterwordspacing

\bibitem{Alexeev_IntelQS}
Y.~Alexeev, ``Evaluation of the {Intel-QS} performance on {T}heta
  supercomputer,'' Argonne National Laboratory - Leadership Computing Facility,
  Technical report {ANL/ALCF} 18/2, Apr 2018.

\bibitem{Haner:2017:PSQ:3126908.3126947}
\BIBentryALTinterwordspacing
T.~H\"{a}ner and D.~S. Steiger, ``0.5 petabyte simulation of a 45-qubit quantum
  circuit,'' in \emph{Proceedings of the International Conference for High
  Performance Computing, Networking, Storage and Analysis}, ser. SC '17.\hskip
  1em plus 0.5em minus 0.4em\relax New York, NY, USA: ACM, 2017, pp.
  33:1--33:10. [Online]. Available:
  \url{http://doi.acm.org/10.1145/3126908.3126947}
\BIBentrySTDinterwordspacing

\bibitem{Boixo_supremacy}
\BIBentryALTinterwordspacing
S.~Boixo, S.~V. Isakov, V.~N. Smelyanskiy, R.~Babbush, N.~Ding, Z.~Jiang, M.~J.
  Bremner, J.~M. Martinis, and H.~Neven, ``Characterizing quantum supremacy in
  near-term devices,'' \emph{Nature Physics}, vol.~14, no.~6, pp. 595--600,
  2018. [Online]. Available: \url{https://doi.org/10.1038/s41567-018-0124-x}
\BIBentrySTDinterwordspacing

\bibitem{Qiskit}
G.~Aleksandrowicz \emph{et~al.}, ``Qiskit: An open-source framework for quantum
  computing,'' 2019.

\bibitem{1601.07195}
\BIBentryALTinterwordspacing
M.~Smelyanskiy, N.~P.~D. Sawaya, and A.~Aspuru-Guzik, ``{qHiPSTER}: The quantum
  high performance software testing environment,'' \emph{arXiv:1601.07195},
  2016. [Online]. Available: \url{https://arxiv.org/abs/1601.07195}
\BIBentrySTDinterwordspacing

\bibitem{Atos}
``{A}tos {Q}uantum {Learning} {Machine},''
  \url{https://atos.net/wp-content/uploads/2018/07/Atos-Quantum-Learning-Machine-brochure.pdf},
  Jun 2018.

\bibitem{ProjectQ}
D.~S. Steiger, T.~H\"{a}ner, and M.~Troyer, ``{ProjectQ}: an open source
  software framework for quantum computing,'' \emph{{Quantum}}, vol.~2, p.~49,
  Jan 2018.

\bibitem{McClean_2016}
\BIBentryALTinterwordspacing
J.~R. McClean, J.~Romero, R.~Babbush, and A.~Aspuru-Guzik, ``The theory of
  variational hybrid quantum-classical algorithms,'' \emph{New Journal of
  Physics}, vol.~18, no.~2, p. 023023, Feb 2016. [Online]. Available:
  \url{https://doi.org/10.1088\%2F1367-2630\%2F18\%2F2\%2F023023}
\BIBentrySTDinterwordspacing

\bibitem{PhysRevLett.110.090501}
\BIBentryALTinterwordspacing
S.~Barrett, K.~Hammerer, S.~Harrison, T.~E. Northup, and T.~J. Osborne,
  ``Simulating quantum fields with cavity {QED},'' \emph{Phys. Rev. Lett.},
  vol. 110, p. 090501, Feb 2013. [Online]. Available:
  \url{https://link.aps.org/doi/10.1103/PhysRevLett.110.090501}
\BIBentrySTDinterwordspacing

\bibitem{Farhi_ApproximateOptimization}
\BIBentryALTinterwordspacing
E.~Farhi, J.~Goldstone, and S.~Gutmann, ``A quantum approximate optimization
  algorithm,'' \emph{arXiv:1411.4028}, Nov 2014. [Online]. Available:
  \url{https://arxiv.org/abs/1411.4028}
\BIBentrySTDinterwordspacing

\bibitem{PhysRevA.92.042303}
\BIBentryALTinterwordspacing
D.~Wecker, M.~B. Hastings, and M.~Troyer, ``Progress towards practical quantum
  variational algorithms,'' \emph{Phys. Rev. A}, vol.~92, p. 042303, Oct 2015.
  [Online]. Available:
  \url{https://link.aps.org/doi/10.1103/PhysRevA.92.042303}
\BIBentrySTDinterwordspacing

\bibitem{1812.07589}
\BIBentryALTinterwordspacing
G.~G. Guerreschi and A.~Y. Matsuura, ``{QAOA} for max-cut requires hundreds of
  qubits for quantum speed-up,'' \emph{arXiv:1812.07589}, Dec 2018. [Online].
  Available: \url{https://arxiv.org/abs/1812.07589}
\BIBentrySTDinterwordspacing

\bibitem{Chen2019}
\BIBentryALTinterwordspacing
Y.~Chen, M.~Farahzad, S.~Yoo, and T.-C. Wei, ``{Detector tomography on IBM
  quantum computers and mitigation of an imperfect measurement},''
  \emph{Physical Review A}, vol. 100, no.~5, p. 052315, Nov. 2019. [Online].
  Available: \url{https://link.aps.org/doi/10.1103/PhysRevA.100.052315}
\BIBentrySTDinterwordspacing

\bibitem{Sarovar2019}
\BIBentryALTinterwordspacing
M.~Sarovar, T.~Proctor, K.~Rudinger, K.~Young, E.~Nielsen, and R.~Blume-Kohout,
  ``{Detecting crosstalk errors in quantum information processors},''
  \emph{Quantum}, vol.~4, p. 321, Sep. 2020. [Online]. Available:
  \url{https://quantum-journal.org/papers/q-2020-09-11-321/}
\BIBentrySTDinterwordspacing

\bibitem{Paladino2014}
\BIBentryALTinterwordspacing
E.~Paladino, Y.~Galperin, G.~Falci, and B.~L. Altshuler, ``{1/ f noise:
  Implications for solid-state quantum information},'' \emph{Reviews of Modern
  Physics}, vol.~86, no.~2, pp. 361--418, 2014. [Online]. Available:
  \url{https://doi.org/10.1103/RevModPhys.86.361}
\BIBentrySTDinterwordspacing

\bibitem{Kjaergaard2020}
\BIBentryALTinterwordspacing
M.~Kjaergaard, M.~E. Schwartz, J.~Braum{\"{u}}ller, P.~Krantz, J.~I.-J. Wang,
  S.~Gustavsson, and W.~D. Oliver, ``{Superconducting Qubits: Current State of
  Play},'' \emph{Annual Review of Condensed Matter Physics}, vol.~11, no.~1,
  pp. 031\,119--050\,605, Mar. 2020. [Online]. Available:
  \url{https://www.annualreviews.org/doi/10.1146/annurev-conmatphys-031119-050605}
\BIBentrySTDinterwordspacing

\bibitem{Heinsoo2018}
\BIBentryALTinterwordspacing
J.~Heinsoo, C.~K. Andersen, A.~Remm, S.~Krinner, T.~Walter, Y.~Salath{\'{e}},
  S.~Gasparinetti, J.-C. Besse, A.~Poto{\v{c}}nik, A.~Wallraff, and C.~Eichler,
  ``{Rapid High-fidelity Multiplexed Readout of Superconducting Qubits},''
  \emph{Physical Review Applied}, vol.~10, no.~3, p. 034040, sep 2018.
  [Online]. Available:
  \url{https://link.aps.org/doi/10.1103/PhysRevApplied.10.034040}
\BIBentrySTDinterwordspacing

\bibitem{IBMQX}
``Ibm quantum experience website,'' \url{https://quantum-computing.ibm.com/},
  accessed: 2020-03-05.

\bibitem{Arute2019}
\BIBentryALTinterwordspacing
F.~Arute, K.~Arya, R.~Babbush, D.~Bacon, J.~C. Bardin, R.~Barends, R.~Biswas,
  S.~Boixo, F.~G. S.~L. Brandao, D.~A. Buell, B.~Burkett, Y.~Chen, Z.~Chen,
  B.~Chiaro, R.~Collins, W.~Courtney, A.~Dunsworth, E.~Farhi, B.~Foxen,
  A.~Fowler, C.~Gidney, M.~Giustina, R.~Graff, K.~Guerin, S.~Habegger, M.~P.
  Harrigan, M.~J. Hartmann, A.~Ho, M.~Hoffmann, T.~Huang, T.~S. Humble, S.~V.
  Isakov, E.~Jeffrey, Z.~Jiang, D.~Kafri, K.~Kechedzhi, J.~Kelly, P.~V. Klimov,
  S.~Knysh, A.~Korotkov, F.~Kostritsa, D.~Landhuis, M.~Lindmark, E.~Lucero,
  D.~Lyakh, S.~Mandr{\`{a}}, J.~R. McClean, M.~McEwen, A.~Megrant, X.~Mi,
  K.~Michielsen, M.~Mohseni, J.~Mutus, O.~Naaman, M.~Neeley, C.~Neill, M.~Y.
  Niu, E.~Ostby, A.~Petukhov, J.~C. Platt, C.~Quintana, E.~G. Rieffel,
  P.~Roushan, N.~C. Rubin, D.~Sank, K.~J. Satzinger, V.~Smelyanskiy, K.~J.
  Sung, M.~D. Trevithick, A.~Vainsencher, B.~Villalonga, T.~White, Z.~J. Yao,
  P.~Yeh, A.~Zalcman, H.~Neven, and J.~M. Martinis, ``{Quantum supremacy using
  a programmable superconducting processor},'' \emph{Nature}, vol. 574, no.
  7779, pp. 505--510, Oct. 2019. [Online]. Available:
  \url{http://dx.doi.org/10.1038/s41586-019-1666-5}
\BIBentrySTDinterwordspacing

\bibitem{RigettiWebSite}
``Rigetti computing website,'' \url{https://www.rigetti.com/what}, accessed:
  2020-11-23.

\bibitem{Schollwoeck2011}
\BIBentryALTinterwordspacing
U.~Schollw{\"{o}}ck, ``{The density-matrix renormalization group in the age of
  matrix product states},'' \emph{Annals of Physics}, vol. 326, no.~1, pp.
  96--192, Aug. 2011. [Online]. Available:
  \url{http://dx.doi.org/10.1016/j.aop.2010.09.012}
\BIBentrySTDinterwordspacing

\bibitem{Orus2014}
\BIBentryALTinterwordspacing
R.~Orus, ``{A Practical Introduction to Tensor Networks: Matrix Product States
  and Projected Entangled Pair States},'' \emph{Annals of Physics}, vol. 349,
  pp. 117--158, Jun. 2013. [Online]. Available:
  \url{http://dx.doi.org/10.1016/j.aop.2014.06.013}
\BIBentrySTDinterwordspacing

\bibitem{markov2008simulating}
\BIBentryALTinterwordspacing
I.~L. Markov and Y.~Shi, ``Simulating quantum computation by contracting tensor
  networks,'' \emph{SIAM Journal on Computing}, vol.~38, no.~3, pp. 963--981,
  2008. [Online]. Available: \url{https://doi.org/10.1137/050644756}
\BIBentrySTDinterwordspacing

\bibitem{villalonga2019flexible}
\BIBentryALTinterwordspacing
B.~Villalonga, S.~Boixo, B.~Nelson, C.~Henze, E.~Rieffel, R.~Biswas, and
  S.~Mandr{\`a}, ``A flexible high-performance simulator for verifying and
  benchmarking quantum circuits implemented on real hardware,'' \emph{npj
  Quantum Information}, vol.~5, pp. 1--16, 2019. [Online]. Available:
  \url{https://doi.org/10.1038/s41534-019-0196-1}
\BIBentrySTDinterwordspacing

\bibitem{huang2019alibaba}
\BIBentryALTinterwordspacing
C.~Huang, M.~Szegedy, F.~Zhang, X.~Gao, J.~Chen, and Y.~Shi, ``Alibaba cloud
  quantum development platform: Applications to quantum algorithm design,''
  \emph{arXiv preprint arXiv:1909.02559}, 2019. [Online]. Available:
  \url{https://arxiv.org/abs/1909.02559}
\BIBentrySTDinterwordspacing

\bibitem{gray2018quimb}
J.~Gray, ``quimb: A python package for quantum information and many-body
  calculations,'' \emph{Journal of Open Source Software}, vol.~3, no.~29, p.
  819, 2018.

\bibitem{QTensor}
D.~Lykov, C.~Ibrahim, A.~Galda, and Y.~Alexeev, ``Tensor network simulator
  {QTensor},'' \url{https://github.com/danlkv/QTensor}, 2020.

\bibitem{wu2019full}
\BIBentryALTinterwordspacing
X.-C. Wu, S.~Di, E.~M. Dasgupta, F.~Cappello, H.~Finkel, Y.~Alexeev, and F.~T.
  Chong, ``Full-state quantum circuit simulationby using data compression,'' in
  \emph{Proceedings of the High Performance Computing,Networking, Storage and
  Analysis International Conference (SC19)}.\hskip 1em plus 0.5em minus
  0.4em\relax Denver, CO, USA: IEEE Computer Society, 2019. [Online].
  Available: \url{https://doi.org/10.1145/3295500.3356155}
\BIBentrySTDinterwordspacing

\bibitem{boixo2017simulation}
S.~Boixo, S.~V. Isakov, V.~N. Smelyanskiy, and H.~Neven, ``Simulation of
  low-depth quantum circuits as complex undirected graphical models,''
  \emph{arXiv preprint arXiv:1712.05384}, 2017.

\bibitem{schutski2019adaptive}
\BIBentryALTinterwordspacing
R.~Schutski, D.~Lykov, and I.~Oseledets, ``An adaptive algorithm for quantum
  circuit simulation,'' \emph{arXiv preprint arXiv:1911.12242}, 2019. [Online].
  Available: \url{https://arxiv.org/pdf/1911.12242.pdf}
\BIBentrySTDinterwordspacing

\bibitem{saleem2020approaches}
\BIBentryALTinterwordspacing
Z.~H. Saleem, B.~Tariq, and M.~Suchara, ``Approaches to constrained quantum
  approximate optimization,'' \emph{arXiv preprint arXiv:2010.06660}, 2020.
  [Online]. Available: \url{https://arxiv.org/abs/2010.06660}
\BIBentrySTDinterwordspacing

\end{thebibliography}

\end{document}